\documentclass{article}

\usepackage{arxiv}
\usepackage[utf8]{inputenc} 
\usepackage{newtxtext}
\usepackage[T1]{fontenc}    
\usepackage{hyperref}       
\usepackage{url}            
\usepackage{booktabs}       
\usepackage{amsfonts}       
\usepackage{amssymb}
\usepackage{amsmath}
\usepackage{float} 
\usepackage{nicefrac}       
\usepackage{microtype}      
\usepackage{cleveref}       
\usepackage{lipsum}         
\usepackage{graphicx}
\usepackage{doi}
\usepackage{pgfplots}
\usepackage{subcaption} 
\usepackage{caption}
\usepackage{dirtree}
\usepackage[english]{babel}
\usepackage{listings}
\usepackage{etoolbox}
\usepackage{csvsimple-l3}
\usepackage{tabularray,siunitx,xfp}
\usepackage{tabularx}
\usepackage{array}
\usepackage{tabularray}
\usepackage{enumitem}
\usepackage{tcolorbox}
\usepackage{csquotes}
\usepackage{adjustbox}

\newcolumntype{P}[1]{>{\centering\arraybackslash}p{#1}}
\pgfplotsset{compat=1.18}
\DeclareGraphicsExtensions{.pdf,.png,.jpg,.svg}

\definecolor{lightyellow}{rgb}{1.0, 0.97, 0.86} 
\definecolor{mangotango}{rgb}{1.0, 0.51, 0.26}
\definecolor{bluebell}{rgb}{0.64, 0.64, 0.82}

\DeclareCaptionFormat{custom}
{%
    \textbf{#1#2}\small #3%
}
\tcbuselibrary{most}
\tcbset{
    codestyle/.style={
        enhanced,
        fontupper=\normalsize\ttfamily\bfseries,
        nobeforeafter,
        tcbox raise base,
        boxrule=0.1pt,
        top=0pt,
        bottom=0pt,
        right=0pt,
        left=0pt,
        arc=3pt,
        boxsep=1.2pt,
        colback=gray!10,
        colframe=gray!50,
        breakable,
       hbox,
        on line,
        valign=center
    }
}

\newcounter{subsubsubsection}[subsubsection] 

\newcommand{\code}[1]{\tcbox[codestyle]{\lstinline!#1!}}

\usepackage{authblk}

\newbox{\orcid}\sbox{\orcid}{\includegraphics[scale=0.06]{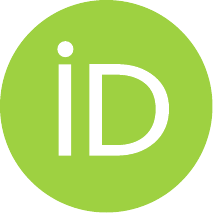}}
\author[1]{\href{https://orcid.org/0000-0003-2867-1706}{\usebox{\orcid}\hspace{1mm}Logan Lang}\thanks{Corresponding author: \texttt{lllang@mix.wvu.edu}}}

\author[2]{\href{https://orcid.org/0000-0002-1164-2856}{\usebox{\orcid}\hspace{1mm}Eduardo Hernandez}\thanks{\texttt{Eduardo.Hernandez@csic.es}}}
\author[3]{\href{https://orcid.org/0000-0001-9737-8074}{\usebox{\orcid}\hspace{1mm}Kamal Choudhary}\thanks{\texttt{kamal.choudhary@nist.gov}}}
\author[1]{\href{https://orcid.org/0000-0001-5968-0571}{\usebox{\orcid}\hspace{1mm}Aldo H. Romero}\thanks{\texttt{Aldo.Romero@mail.wvu.edu}}}

\affil[1]{Department of Physics, West Virginia University, Morgantown, WV 26506, United States}
\affil[2]{Instituto de Ciencia de Materiales de Madrid, Campus de Cantoblanco, C. Sor Juana Inés de la Cruz, 3, Fuencarral-El Pardo, Madrid 28049, Spain}
\affil[3]{National Institute of Standards and Technology, 100 Bureau Dr, Gaithersburg, MD 20899, United States}

\title{ParquetDB: A Lightweight Python Parquet-Based Database}

\begin{document}

\maketitle

\begin{abstract}
Traditional data storage formats and databases often introduce complexities and inefficiencies that hinder rapid iteration and adaptability. To address these challenges, we introduce ParquetDB, a Python-based database framework that leverages the Parquet file format's optimized columnar storage. ParquetDB offers efficient serialization and deserialization, native support for complex and nested data types, reduced dependency on indexing through predicate pushdown filtering, and enhanced portability due to its file-based storage system. Benchmarks show that ParquetDB outperforms traditional databases like SQLite and MongoDB in managing large volumes of data, especially when using data formats compatible with PyArrow. We validate ParquetDB's practical utility by applying it to the Alexandria 3D Materials Database, efficiently handling approximately 4.8 million complex and nested records. By addressing the inherent limitations of existing data storage systems and continuously evolving to meet future demands, ParquetDB has the potential to significantly streamline data management processes and accelerate research development in data-driven fields.
\end{abstract}


\section{Introduction}

In an era where data has become the lifeblood of technological innovation, the need for efficient, scalable, and accessible storage solutions is more important than ever. As we push the boundaries of artificial intelligence and data-driven decision-making, our tools for managing data must adapt to keep pace with these rapid changes. Traditional data storage formats such as CSV, JSON, and TXT, along with database management systems like SQLite \cite{SQLite} and MongoDB\cite{MongoDBDeveloperData}, have long served as the backbone for data handling in various applications \cite{habyarimanaGenomicsData2021, jainCommentaryMaterialsProject2013, WellKnownUsersSQLite, CustomerCaseStudies}. While these formats and systems are powerful and offer certain advantages, they often introduce complexities and inefficiencies that hinder quick iteration and experimentation; elements that are crucial for cutting-edge research and development. These complexities are reflected in the coding implementations that demand significant effort to optimize data retrieval, indexing, and storage efficiency. Moreover, while traditional systems may offer robust features, their rigidity often complicates adaptation to new data structures or evolving project needs. This rigidity creates friction in a field that thrives on adaptability and quick responses to new challenges \cite{YouUseRight,EvolutionDataManagement,iabacrEvolutionDataEngineering2023,naseemEvolutionLargeScaleData2024}.

Traditional data storage formats, such as JSON and CSV files, are popular choices due to their simplicity, structured format, and human readability. However, this simplicity comes at a cost, leading to significant performance limitations that stem from their inherent design. JSON and CSV files are ASCII/UTF encoded, which is often referred to as plain text encoding. Encoding involves converting data from one format to another. To store text in Random Access Memory (RAM), data must be represented as bytes or sequences of 1s and 0s (bits). ASCII/UTF provides a standard on how each character in a text file is mapped into bytes; for example, the character "0" is represented as 00110000 in binary.

\begin{figure}[t]
    \centering
    \includegraphics[width=\textwidth]{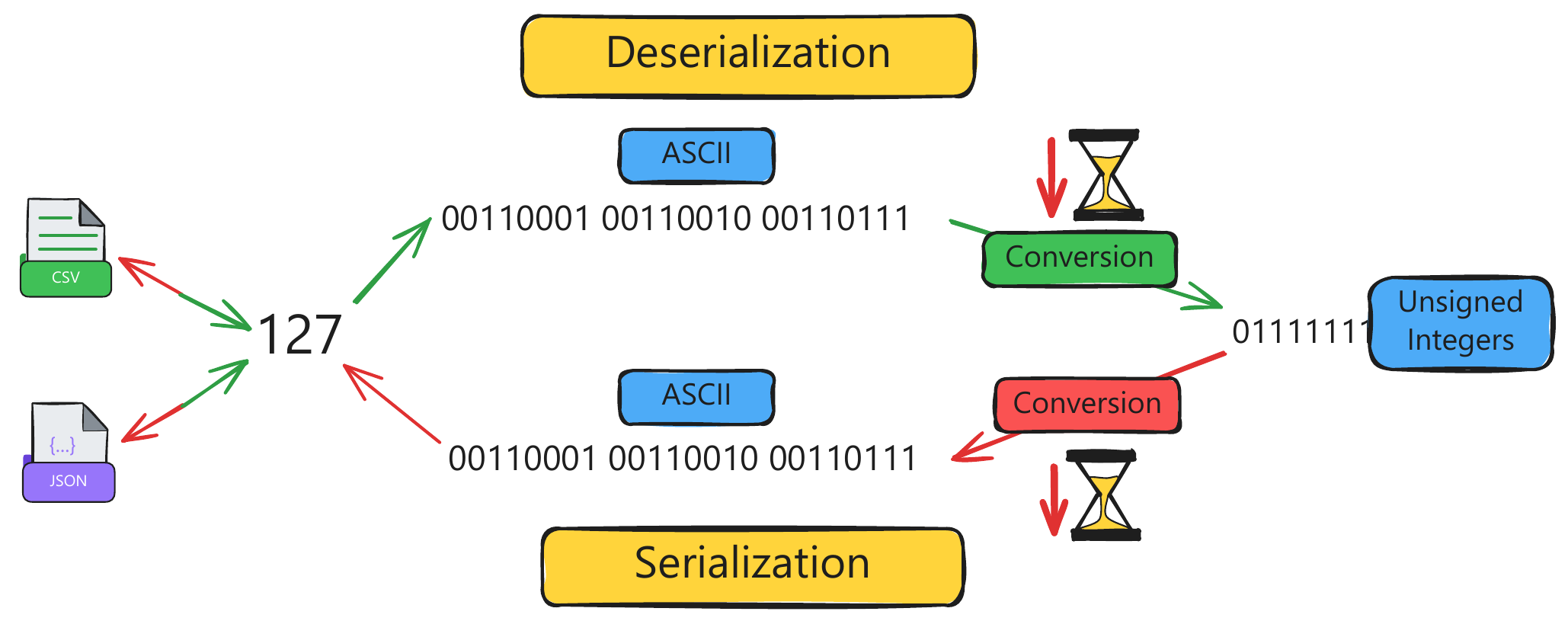} 
    \caption{ Illustration of the serialization and deserialization process for numerical data in CSV and JSON formats. The inefficient ASCII encoding requires multiple bytes for each character, leading to larger file sizes and slower data conversion. Conversion steps, represented by hourglass icons, introduce additional latency as numerical values must be transformed into optimized binary formats for computational use. This process highlights the performance bottleneck in traditional plain text file formats.}
    \label{fig:serilization-deserialization}
\end{figure}

For non-numerical text data, this encoding is sufficient. However, when the data consists of numerical values like integers (int), floats, or boolean values (True/False), this encoding becomes highly inefficient, as it requires a byte for each character in a number. For instance, the integer 127 requires three bytes (00110001 00110010 00110111) to represent in ASCII, as each digit must be encoded separately. See Figure \ref{fig:serilization-deserialization} for a visual representation of this encoding process By changing the encoding standard for numerical values, we could represent many numbers in fewer bytes. For example, if we use a convention where "0" is represented as 00000000 and "1" as 00000001, then "127" can be encoded as 01111111, which takes only a single byte. Compared to ASCII, this results in a 66\% reduction in storage space. The reduction in space usage also leads to decreased read times from disk to memory.

The issue of inefficient encoding lies at the core of the performance limitations of these file formats. As datasets grow in scale, the corresponding increase in the number of bytes exacerbates file size expansion.  This growth in data volume directly contributes to slower retrieval speeds due to heightened I/O operations and increased processing demands, necessitated by the need to scan larger quantities of data.  Further complicating performance, the serialization and deserialization create well-known bottlenecks within these file formats \cite{caoAcceleratingDataSerialization2022}. These bottlenecks are illustrated in Figure \ref{fig:serilization-deserialization}, highlighting the performance impact of converting data between ASCII and binary formats. For example, the conversion of numerical data from ASCII encoding to integer format for mathematical operations, which comes at a cost. This overhead can substantially impair application responsiveness, thereby hindering the pace of research and development iterations. Addressing the inefficiencies in encoding is thus a critical consideration in the pursuit of more effective data storage solutions. \cite{stephenCSVVsParquet2021,pJSONCSVParquet2023,ApacheParquetVs,engineerCSVVsParquet2022,15SparkFile}.

Traditional database management systems offer a structured solution for data storage, particularly due to their support for optimal encoding, which improves the efficiency of serialization and deserialization processes. Unlike JSON and CSV files, databases support advanced features like data consistency and efficient querying. For instance, they employ automated indexing and have built-in mechanisms to ensure data integrity across updates, making it easier to manage data consistently. Although similar indexing can technically be implemented in JSON and CSV files, this approach demands significant discipline and is only practical for small datasets. As datasets grow in size, managing and querying data in JSON or CSV formats becomes increasingly challenging, especially because they require full scans and additional deserialization for certain data types (e.g., integers, floats, and booleans). This limitation, manageable for small data volumes, becomes impractical with larger datasets due to the substantial time required for processing.

\begin{table}[t]
    \begin{adjustbox}{max width=\textwidth,center}
    \begin{tblr}{
        colspec = {X[0.25\textwidth,c,m]X[0.75\textwidth,c,m]},
        column{1} = {valign=m, halign=c}, 
        column{2} = {halign=l},
        cells = {font=\footnotesize\linespread{0.8}\selectfont},
        stretch = 0,
        rowsep = 3pt,
        hlines = {black, 1pt},
        vlines = {black, 1pt},
      }
    \SetCell[c=2]{c}\textbf{Features and Benefits} \\ 
    \textbf{Simple Interface} & Easy-to-use methods for creating, reading, updating, and deleting data. \\
    \textbf{High Performance} & Utilizes Apache Parquet and PyArrow for efficient data storage and retrieval. \\
    \textbf{Complex Data Types} & Handles nested and complex data types. (Ex. nd arrays, lists, dictionaries, etc.) \\
    \textbf{Portability} &  File-based storage, allows for easy transfer.  \\
    \textbf{Schema} & Contains a schema that describes the data, ensuring consistency. \\
    \textbf{Schema Evolution} & Supports adding new fields and updating schemas. \\
    \textbf{Predicate Pushdown} & Optimizes queries by reading only relevant data blocks. \\
    \textbf{Column Pushdown} & Selects columns to read into memory. \\
    \textbf{Efficient Encoding} & Choice of field-level encoding. \\
    \textbf{Efficient Compression} & Choice of field-level compression. \\
    \textbf{Metadata Support} & Table and field-level metadata support. \\
    \textbf{Batching Support} & Files are grouped to facilitate batching. \\
    \end{tblr}
    \end{adjustbox}
    \caption{Features and benefits of ParquetDB}
    \label{table:features}
\end{table}

However, traditional database systems have their own challenges, often linked to the specific type of database used. Databases fall into two main categories: relational databases (e.g., SQLite) and non-relational (NoSQL) databases, which encompass document-based (e.g., MongoDB), key-value, columnar, and graph databases \cite{RelationalDatabaseVs, WhatAreDifferent}. While these databases offer ease of use, their architectures impose inherent limitations that can hinder performance and scalability, especially with large volumes of unstructured data. For example, relational databases enforce a fixed schema that ensures data integrity but poses challenges when the data model evolves over time. On the other hand, document databases lack strict schema enforcement, which makes them more flexible in handling evolving data models. However, this flexibility also compromises data integrity, making them more susceptible to inconsistently typed data entries. Further limitations, which this paper will explore in depth, include difficulties in managing complex data types and relationships, the need for continuous index tuning to sustain optimal query performance, and challenges in accommodating flexible or evolving data models \cite{singh15TypesDatabases2024}. Additionally, factors like portability (server-based vs. serverless configurations) and the increased code complexity required to integrate these databases into applications add to the overall overhead associated with database management.

To address the limitations of traditional data storage options, we introduce ParquetDB, a Python-based database framework designed to leverage the advantages of the Parquet file format \cite{Parquet}. Parquet is a columnar storage format optimized for efficient serialization and deserialization, improving performance and storage efficiency. The Parquet file format also offers unique features such as table- and column-level metadata, column-level encoding and compression, column-level statistics, schema enforcement, and native batching format. These benefits make ParquetDB a scalable, efficient, and portable option that overcomes the inherent limitations of traditional data storage options. A summary can be of the features can be found in Table \ref{table:features}

In this paper, we discuss the design and implementation of ParquetDB, illustrating how it addresses the challenges posed by traditional data storage options. We evaluate its performance and scalability through benchmarks and demonstrate its applicability in real-world scenarios. By providing a robust and user-friendly alternative, ParquetDB has the potential to significantly streamline data management processes in iterative research and development environments.

\section{Installation and Usage of ParquetDB}

\subsection{Installation}
To install ParquetDB, you can use pip, Python's package installer. Run the following command in your terminal:

\begin{lstlisting}
pip install parquetdb
\end{lstlisting}

Ensure you have Python 3.8 or above installed, as ParquetDB is compatible with these versions.

\subsection{Basic Usage}

Once installed, you can start using ParquetDB in your Python projects. Here is a simple example to get you started:

\begin{lstlisting}
from parquetdb import ParquetDB

# Initialize the database
db = ParquetDB(db_path='parquetdb')

# Create data
data = [
    {'name': 'Alice', 'age': 30, 'occupation': 'Engineer'},
    {'name': 'Bob', 'age': 25, 'occupation': 'Data Scientist'}
]

# Add data to the database
db.create(data)

# Read data from the database
employees = db.read()
print(employees.to_pandas())
\end{lstlisting}

\subsection{Github}

For more details, including advanced features and contributions, please visit the GitHub repository at: \code{https://github.com/lllangWV/ParquetDB}
The repository contains additional examples, API documentation, and guidelines for contributing to the project.

\section{Background on Traditional Data Storage Methods}
In the initial stages of a project, selecting an appropriate data storage format is paramount for facilitating effective data management and subsequent analysis. When project requirements are ambiguous or still evolving, developers often resort to data formats that are easily accessible and human-readable, such as \code{.txt}, \code{.csv}, or \code{.json}. Each of these formats has distinct advantages and inherent limitations. For instance, while a \code{.csv} file is intuitive, human-readable, and adept at handling structured tabular data, it struggles with managing unstructured or hierarchical data. In contrast, \code{.json} is capable of representing more complex, nested data structures, making it a versatile choice for scenarios involving semi-structured or unstructured data. However, using JSON sacrifices the inherent benefits of structured tabular data, such as the ease of defining and enforcing relationships between datasets. Both file formats, nonetheless, exhibit significant limitations when data volumes become substantial, particularly in terms of scalability and efficient data manipulation. Thus, a comprehensive understanding of these limitations and an informed selection of the appropriate storage solution are vital, as storage decisions can have substantial ramifications for downstream performance, scalability, and project adaptability.

In the following sections, we will examine two primary categories of data storage: file-based formats, such as those mentioned above, and database systems, including technologies like \code{SQLite}, \code{MongoDB}, and \code{PostgreSQL}. File-based formats offer a lightweight, flexible approach for managing data, particularly in smaller-scale projects or where simplicity is prioritized. However, despite their flexibility, file-based systems inherently lack mechanisms for ensuring data integrity or the unique management of records. These systems often require auxiliary support from programming logic to achieve adequate data integrity control, making them suboptimal for more complex data environments.

To address these limitations, database systems are employed to provide more reliable and consistent data management. Database systems are designed with built-in capabilities for efficient data querying, relationship enforcement, and scalability. Broadly, there are two primary categories of database systems: relational databases (e.g., \code{MySQL}, \code{PostgreSQL}) that employ structured schemas to manage data in a tabular format, and NoSQL databases (e.g., \code{MongoDB}, \code{Cassandra} \cite{ApacheCassandraApache}) that provide the flexibility necessary for handling unstructured or semi-structured data. Relational databases excel in scenarios requiring strict consistency, well-defined relationships, and complex querying, while NoSQL databases are ideal for applications that need to accommodate dynamic, large-scale, and varied data structures. Understanding the strengths and limitations of both approaches is crucial for making informed data storage decisions that align with the structure, scale, and evolving requirements of a project.

\subsection{File-Based Data Storage Formats}

\begin{table}[t]
    \begin{adjustbox}{width=\textwidth,center}
    \begin{tblr}{
        colspec = {X[0.2\textwidth,c,m]X[0.2\textwidth,c,m]X[0.20\textwidth,c,m]X[0.20\textwidth,c,m]X[0.20\textwidth,c,m]},
        column{1} = {valign=m, halign=c}, 
        stretch = 0,
        rowsep = 3pt,
        hlines = {black, 1pt},
        vlines = {black, 1pt},
      }
    \SetCell[c=5]{c}\textbf{Comparison of File Storage Formats} \\ 
    \textbf{Features} & \textbf{.txt} & \textbf{.csv} & \textbf{.json} & \textbf{.parquet} \\ 
    \textbf{Encoding} & UTF-8 or ASCII & UTF-8 or ASCII & UTF-8 or ASCII & Binary \\ 
    \textbf{Structure} & Unstructured & Tabular (structured) & Hierarchical (semi-structured) & Columnar (structured) \\ 
    \textbf{Human Readability} & High & High & Moderate & Low \\ 
    \textbf{Portability} & High & High & High & High \\ 
    \textbf{Scalability} & Limited & Moderate & Moderate & High \\ 
    \textbf{Batch Support} & No & Yes & No & Yes \\ 
    \textbf{Schema Enforced} & No & No & No & Yes \\ 
    \end{tblr}
    \end{adjustbox}
    \caption{Comparison of File Storage Formats: This table compares different file storage formats (TXT, CSV, JSON, and Parquet) across various features.}
    \label{table:file-storage-comparison}
\end{table}

File-based data storage formats represent a fundamental approach to storing and organizing data in files, without the complexities inherent to a database management system. Although these formats can be supplemented with programming tools to emulate basic database functionalities, such as indexing, querying, and data manipulation, they are often best suited for smaller datasets, prototyping, or scenarios where simplicity and human readability are prioritized. For instance, \code{.csv} files are frequently employed for quickly prototyping data analysis workflows, whereas \code{.json} files are particularly useful for exchanging configuration data between applications. Their single-file structure enhances portability, facilitating straightforward data transfer across different environments. Each format exhibits distinct characteristics that make it appropriate for specific use cases but also carries inherent limitations, particularly as data complexity and scale increase.

In the following sections the \code{.txt}, \code{.csv}, \code{.json} formats will provide a detailed examination of each file-based data storage format, with the exception of \code{.parquet} files, as they are addressed separately due to ParquetDB's reliance on this file type. A summary of the comparison of these formats is available in Table \ref{table:file-storage-comparison}. This table contrasts the formats across several dimensions, including encoding, structure, human readability, portability, scalability, batch support, and schema enforcement.

\subsubsection{Plain Text Files (txt)}
Plain text (TXT) files represent one of the most fundamental and widely utilized formats for storing unstructured data. They consist of raw text devoid of specific formatting or metadata, rendering them highly adaptable for diverse applications. TXT files are typically used for storing configuration settings, logs, documentation, or any information requiring simple storage without the complexity inherent to more sophisticated data models. These files are conventionally saved with the \code{.txt} extension and can be accessed through any basic text editor. TXT files are commonly encoded using either ASCII or UTF-8, with UTF-8 being the preferred encoding due to its ability to support an extensive range of characters, including international scripts, thereby enhancing the versatility of TXT files across linguistic and cultural contexts \cite{iqbalTXTTextDocument2019}.

TXT files are inherently unstructured, lacking any predefined schema, delimiters, or formatting. They store data as raw text, which makes them suitable for simple content that does not require relationships or hierarchy. Unlike formats like CSV or JSON, TXT files do not organize data into structured records, rows, or key-value pairs, which limits their applicability for more complex data storage needs \cite{WhatAreDisadvantages}. However, this simplicity is also an advantage, as TXT files are highly readable and can be accessed and understood without specialized software. Their straightforward format makes them ideal for storing logs, notes, and configuration files, adding to their practicality .

Despite their advantages, TXT files have notable limitations in terms of scalability. As the size of the data grows, TXT files become increasingly challenging to manage due to the absence of indexing or efficient search capabilities \cite{WhatAreDisadvantages}. Operations like searching for specific information or editing large TXT files can be cumbersome and slow, making scalability a significant issue when dealing with large-scale data storage. Furthermore, TXT files do not offer any inherent data validation mechanisms. Since they store raw text without any structured schema or constraints, there is no way to enforce data integrity or validate the correctness of the content, making them prone to errors, especially when manually edited. Applications that process TXT files must implement their own validation procedures to ensure data accuracy.

While TXT files can support batch processing to some extent, as they can be read and processed line by line, the lack of structure means that additional parsing logic is often required to extract meaningful information. Many programming languages can easily process TXT files, but the absence of a formal structure can make batch operations less efficient compared to structured formats. Nonetheless, TXT files remain universally compatible, supported by all major operating systems, applications, and text editors. The lack of specialized formatting and the use of standard encodings make TXT files accessible to a wide range of users, regardless of their technical expertise.

Another key characteristic of TXT files is that they are inherently schema-less, meaning they have no predefined structure or rules governing their content. This lack of schema enforcement allows for maximum flexibility in what can be stored, but it also means that there is no consistency or integrity guaranteed in the data. Users are free to store any kind of information without restrictions, which can lead to inconsistencies and errors if the data is not carefully managed. This flexibility is advantageous for simple, ad-hoc storage needs but limits the reliability of TXT files for more complex data management tasks.

\subsubsection{Comma-Separated Values Files (.csv)}

Comma-Separated Values (CSV) files constitute one of the most widely adopted formats for the storage of tabular data. Each CSV file represents data in a row-by-row format, with individual fields within a record separated by commas. Each line in a CSV file corresponds to a distinct record, and the use of a comma as a delimiter separates each field within that record. CSV is inherently a plain-text format, frequently utilized for exporting data from databases, spreadsheets, or other applications due to its straightforward nature and ease of integration. Typically, CSV files are encoded using ASCII or UTF-8, with UTF-8 being the default choice for many implementations. \cite{CSVFormatHistory2021,CSVCommaSeparated,shafranovichCommonFormatMIME2005}.

The structure of CSV files is fundamentally simple, consisting of rows of text where each field is separated by a comma delimiter. Generally, the first row serves as a header, providing labels for each field. This straightforward tabular format makes CSV files particularly well-suited for representing flat, two-dimensional data, but it limits their capability to represent more complex or hierarchical data structures. The absence of inherent nesting features confines CSV files to datasets with uniform fields across records. While the simplicity of CSV files contributes to their human readability, particularly when opened in spreadsheet applications such as Microsoft Excel or Google Sheets, readability can diminish significantly for very large datasets or when data fields contain commas, necessitating careful use of escape characters and formatting conventions.

Despite the advantages, CSV files exhibit notable limitations in terms of scalability. As datasets grow in size, managing CSV files becomes increasingly impractical \cite{WhatAreChallenges}. The absence of built-in indexing mechanisms results in inefficiencies when performing search or retrieval operations on large CSV files, making such processes resource-intensive. Additionally, as the number of rows or columns increases, the text-based nature of CSV files can lead to considerable inefficiencies in terms of both storage requirements and performance. CSV files also lack intrinsic data validation mechanisms, offering no means to enforce data types or maintain data integrity. Consequently, errors such as missing values or incorrect formats are prevalent, particularly when data is manually edited. To mitigate these issues, applications that process CSV files must implement their own validation rules to ensure data consistency and accuracy.

However, CSV files retain their utility for batch processing operations. Owing to their line-by-line structure, CSV files can be efficiently read and processed in chunks, which is conducive to batch data operations \cite{yanchevMostTimeEfficient2020}. Numerous programming languages provide robust libraries that support the streaming and processing of CSV data in a memory-efficient manner, enabling the use of moderately large datasets in batch-oriented tasks. CSV files are also characterized by their universal compatibility \cite{HowEfficientlyRead2023}. They are supported by virtually all database systems, spreadsheet software, and data analysis tools. The pervasive use of CSV files ensures that they are often the default format for data import and export, facilitating seamless data exchange across disparate systems. Their compatibility across different platforms makes CSV files a reliable choice for data sharing and collaboration in diverse environments.

CSV files are inherently schema-less, lacking a formal definition of data types, column constraints, or inter-record relationships. This absence of an enforced schema allows users to implicitly define data structure, often through documentation or the use of column headers. While the flexibility afforded by a schema-less format can be advantageous in rapidly evolving contexts, it also introduces the risk of inconsistencies if different users or applications interpret or modify the data without adhering to a consistent structure. The lack of schema enforcement can thus lead to ambiguities, undermining the reliability of CSV files for complex data management requirements \cite{WhyYouDont2019}.

\subsubsection{JavaScript Object Notation (.json)}

JSON (JavaScript Object Notation) is a widely adopted data interchange format, predominantly employed in web development and APIs for data storage and exchange \cite{JSON}. JSON files represent data as key-value pairs and support hierarchical, nested structures, making them well-suited for modeling complex data relationships. This format strikes a balance between human readability and machine parsability, providing an efficient solution for serializing structured and semi-structured data. By default, JSON files are encoded using UTF-8, which supports a broad spectrum of characters, including international symbols, thereby ensuring compatibility across diverse systems and programming environments.

JSON is considered a semi-structured data format that adeptly represents hierarchical information. It accommodates a variety of data types, including strings, numbers, booleans, arrays, and objects. JSON's capacity to handle nested objects makes it particularly advantageous for representing complex and relational data, such as configurations or multi-layered settings. One of the notable strengths of JSON is its human-readable syntax, designed for ease of comprehension, which facilitates straightforward inspection and debugging by developers. Unlike binary formats, JSON is stored as plain text with a well-defined structure that employs brackets, rendering it accessible without the need for specialized tools.

Despite its strengths, JSON poses challenges in terms of scalability. Its text-based nature results in inefficiencies when managing very large datasets, leading to increased file sizes and slower parsing times. Moreover, JSON lacks intrinsic indexing capabilities, making search operations within extensive JSON files less efficient compared to more sophisticated data storage systems. JSON also does not inherently enforce data validation. While schema definitions, such as JSON Schema, can be used to impose validation rules, JSON itself lacks native mechanisms for ensuring data integrity. This can lead to inconsistencies, particularly if data is improperly formatted or altered outside of controlled environments.

Furthermore, JSON files do not natively support batch processing capabilities. Handling multiple records or datasets often requires loading the entire JSON file into memory, which can be resource-intensive for large datasets. To manage batch operations effectively, dedicated tools or custom scripts are typically necessary. Nevertheless, JSON's compatibility with a wide range of systems and applications is a major advantage. As a standard format for data interchange, it enjoys comprehensive support across all major programming languages, making it an excellent choice for data exchange between heterogeneous systems. This broad compatibility has contributed significantly to JSON's widespread use in web development and data APIs.

JSON is inherently schema-less, meaning it does not necessitate a predefined structure, which affords flexibility in data representation. This characteristic facilitates rapid prototyping and iterative development of data models. However, the absence of an enforced schema can also lead to inconsistencies, particularly if the data model evolves frequently or if different components of an application require distinct data structures. JSON Schema can be utilized to introduce structure and validation, but such enforcement is external and not intrinsic to the JSON format itself.

\subsection{Traditional Database Systems}

Database Management Systems (DBMS) are sophisticated software solutions designed to efficiently manage, store, and retrieve data. They provide a structured approach to data management, which is essential for applications that demand robust data integrity, complex querying capabilities, and high scalability. Fundamentally, a database management system must support CRUD operations—Create, Read, Update, and Delete. These operations allow users and applications to insert new records, retrieve existing information, modify data, and delete records as needed. The effective implementation of CRUD operations is critical to ensuring data consistency and reliability, which are foundational to maintaining the integrity of information systems. DBMS play a pivotal role in enabling data to be accessed, modified, and controlled efficiently, thereby serving as the backbone of a wide array of modern software applications.

The following subsections will explore key concepts of database management systems, such as ACID transactions, indexing mechanisms, types of databases, and the distinction between server-based and serverless databases. We will also discuss MongoDB and SQLite, two distinct types of database management systems that exemplify the differences between document-based and relational approaches to data management. These systems will later be used to benchmark ParquetDB, providing insight into their comparative performance. Table \ref{table:database-comparison} summarizes the features of SQLite, MongoDB, and ParquetDB, highlighting key aspects such as ACID compliance, concurrency, portability, scalability, access methods, native support for complex data structures, schema evolvability, encoding techniques, batch support, and indexing methods.

\subsubsection{Atomicity, Consistency, Isolation, and Durability (ACID)}

ACID transactions are a foundational concept in database management systems (DBMSs) that ensure data reliability and integrity during operations. The acronym ACID stands for \code{Atomicity}, \code{Consistency}, \code{Isolation}, and \code{Durability}—four essential properties that collectively uphold the correctness and stability of transactions, even amidst system failures, concurrent operations, or unexpected errors. These properties are critical for maintaining trust in databases deployed in mission-critical applications, such as financial systems, healthcare, and enterprise resource planning.

\code{Atomicity} guarantees that a transaction is executed as an all-or-nothing operation, meaning that either every operation within the transaction is completed successfully, or none are applied. This property is crucial for preventing partial updates that can lead to data inconsistencies. For example, in a banking application, a fund transfer must either deduct the amount from one account and add it to another in its entirety or not occur at all. This ensures no partial transactions that could compromise the accuracy of the financial data.

\code{Consistency} ensures that the database transitions from one valid state to another before and after a transaction. Each transaction must adhere to all predefined rules, constraints, and relationships, ensuring that the database remains in a consistent state. This property is particularly vital in applications with complex and interdependent data relationships, as it prevents violations of integrity rules that could jeopardize data reliability.

\code{Isolation} guarantees that concurrent transactions are executed independently without interference, ensuring that the execution of one transaction does not impact the intermediate state of another. This property is critical in high-transaction-volume environments, where managing concurrent data access is necessary to prevent conflicts and maintain data accuracy. Isolation preserves the illusion that each transaction occurs in isolation, which is particularly important for applications requiring high levels of concurrent processing.

\code{Durability} ensures that once a transaction has been committed, the changes are permanent and will persist even in the event of a system failure. This property means that the results of a successful transaction are recorded reliably, ensuring resilience against power outages, crashes, or other types of disruptions. Durability is of utmost importance in applications where data loss is unacceptable, such as those managing financial records or medical information systems.

\subsubsection{Indexing and why it is important}

Database indexing is a critical optimization technique employed to enhance the efficiency of data retrieval operations within a database. By constructing auxiliary data structures, indexes enable faster access to specific records, significantly reducing the time complexity of query execution. Indexes serve as an efficient lookup mechanism, allowing the database to rapidly locate the desired rows without the need to perform exhaustive scans of entire tables. This makes indexing indispensable for maintaining performant large-scale databases, particularly when executing complex queries involving sorting, filtering, or joins.

The importance of indexing in databases cannot be overstated \cite{WhatDatabaseIndex}. In the absence of indexes, a database must resort to full table scans for each query, leading to substantial performance degradation as the dataset size increases. Indexes dramatically reduce query response times, thereby enhancing the overall performance of applications that demand rapid data access, such as e-commerce platforms, financial systems, and high-frequency trading applications. Furthermore, indexes are essential for enforcing data integrity constraints, such as unique keys, by preventing duplicate values and ensuring data consistency.

\begin{figure}[t]
    \centering
    \includegraphics[width=\textwidth]{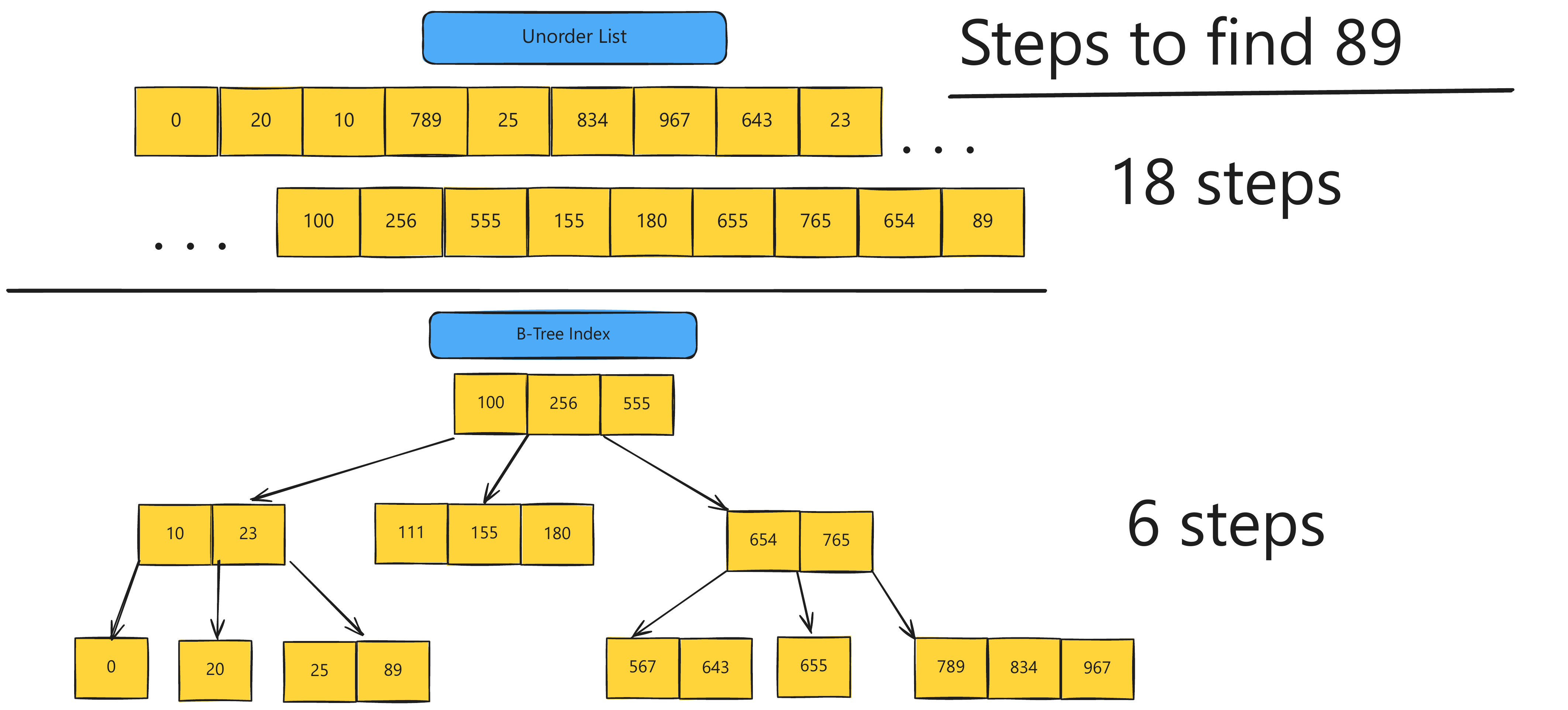}
    \caption{Comparison of search efficiency in an unordered list versus a B-Tree index. The unordered list requires 18 steps to locate the value 89 through a sequential scan, while the B-Tree index reduces this to only 6 steps by leveraging its hierarchical and balanced structure. This illustrates the significant performance improvement that indexing provides, particularly in large datasets, by minimizing the number of operations needed for data retrieval.}
    \label{fig:importance-of-indexing}
\end{figure}

There are several types of indexing structures utilized in databases, each suited to different types of queries and data characteristics. The most common indexing structures include:

\code{B-Tree Indexes}: B-Tree (Balanced Tree) indexes are among the most widely utilized indexing structures in relational databases. A B-Tree is a self-balancing tree data structure that maintains data in a sorted order, facilitating efficient insertion, deletion, and search operations. Each node in a B-Tree can contain multiple keys and children, which ensures that the tree remains balanced, thereby keeping the height of the tree minimal. This minimal height is crucial for reducing the number of disk accesses required during search operations, ultimately enhancing performance. As illustrated in Figure \ref{fig:importance-of-indexing}, a B-Tree index significantly reduces the number of steps required to locate a specific value compared to an unordered list. B-Trees are highly effective for indexing because they provide predictable performance for both read and write operations, and their structure is well-suited for managing large volumes of data in secondary storage. It is worth noting that B-Trees are a generalization of binary search trees, providing enhanced balance and efficiency in managing large datasets \cite{shandilyaBTreeIndexingBasics2024}.

\code{Binary Search Tree Indexes}: Binary Search Trees (BSTs) represent a simpler form of tree structure used for indexing, in which each node has at most two children—the left and right child. The left subtree of a node contains values less than the node's key, while the right subtree contains values greater than the node's key. Although BSTs can be effective for indexing smaller datasets with balanced distributions, they are prone to becoming unbalanced, which can lead to degraded performance in worst-case scenarios. B-Trees were developed to address these limitations by maintaining a balanced structure, making them more suitable for large-scale databases \cite{BinarySearchTree}.

\code{Hash Indexes}: Hash indexes employ hash functions to map keys to specific locations in a table. This type of indexing is highly efficient for equality searches, where the objective is to find records that match a specific value. However, hash indexes are not well-suited for range queries, as the hash function does not preserve any order of the indexed values, thereby limiting their applicability in certain query types \cite{patidarHashIndexing2024}.

\code{Bitmap Indexes}: Bitmap indexes are typically employed for columns with a small number of distinct values, such as categorical fields like gender or status. Bitmap indexes utilize bit arrays (bitmaps) to indicate the presence or absence of specific values, which can result in significant storage savings and rapid query processing for particular analytical workloads. This makes bitmap indexing especially advantageous in data warehousing and business intelligence applications \cite{HowBitmapIndex2017}.

While indexing provides substantial performance benefits, it is not without cost \cite{cobbIndexBasicsHidden2017}. The primary cost associated with indexing is the additional storage required to maintain these index structures. Depending on the number of indexes and the type of indexing utilized, the storage overhead can be considerable. Moreover, indexes must be updated whenever the underlying data is modified, which introduces additional computational overhead during insert, update, or delete operations. This trade-off between improved read performance and increased write costs must be carefully evaluated, particularly for write-intensive applications where maintaining multiple indexes may significantly hinder performance.

\subsubsection{Server-based, Classic Serverless, and Neo/Cloud-based serverless}

Databases can be broadly classified into three primary types based on their deployment and management models: server-based databases, classic serverless databases, and neo-serverless (cloud serverless) databases \cite{SQLiteServerless}. These three paradigms offer distinct advantages and limitations, making them suitable for different use cases depending on organizational requirements and application demands.

\textbf{Server-Based Databases:}  Server-based databases are the traditional model for database management, requiring dedicated server infrastructure for hosting and maintenance \cite{DatabaseServerOverview}. An example of server based databases are Postgres and MongoDB, these databases run on servers, either on-premises , and are managed by database administrators (DBAs) who handle tasks like configuration, maintenance, scaling, security, and backups. The operational overhead of server-based databases is significant, involving hardware provisioning, software installation and configuration, performance tuning, and ensuring system availability through constant monitoring and maintenance.

Server-based databases continue to be popular due to their established role in enterprise environments. They offer complete control over database configuration, security policies, and performance optimization, making them appealing for organizations needing fine-grained control over their data infrastructure. Additionally, server-based databases are ideal for applications with predictable workloads, as dedicated resources can be optimized for consistent performance. This model provides extensive control over database configurations, performance tuning, and custom setups, making it suitable for applications that need a high degree of customization and direct management.

\textbf{Classic Databases:} Classic serverless databases, in the context of this discussion, refer to lightweight databases such as SQLite that do not require dedicated server infrastructure. Unlike server-based databases, SQLite is a self-contained, embedded database engine that operates within the application itself, thus obviating the need for any external server configuration or management \cite{SQLiteServerless}. Users interact directly with the database through the application, eliminating the necessity of managing server infrastructure. Serverless databases like SQLite are designed to provide simplicity and ease of use, particularly for applications where the overhead associated with server setup and maintenance is unnecessary. By embedding the database within the application, SQLite allows developers to integrate a data storage solution with minimal configuration, making it particularly well-suited for prototyping, small-scale applications, and embedded systems. Since SQLite operates as an embedded component, it provides a lightweight and self-sufficient solution without requiring a dedicated DBA.

\textbf{Neo-Serverless (Cloud Serverless)}: The database engine runs in a separate namespace from the application, probably on a separate machine, but the database is provided as a turn-key service by the hosting provider. It requires no management or administration by the application owners, and is so easy to use that the developers can think of the database as being serverless even if it really does use a server under the covers \cite{WhatCloudDatabase2023, WhatCloudDatabase}. Microsoft Azure Cosmos DB \cite{ealsurServerlessDatabaseComputing2024} and Amazon S3 \cite{AmazonS3Cloud} are examples of neo-serverless databases. These databases are implemented by server processes running separately in the cloud. But the servers are maintained and administered by the ISP, not by the application developer. Application developers just use the service without provisioning, configuring, or managing database server instances, as all of that work is handled automatically by the service provider.

\vspace{0.2cm}
\textbf{Comparative Benefits}

\begin{itemize}[label=-]
    \item \textbf{Control vs. Simplicity:}  Server-based databases provide extensive control over hardware, software, and configuration settings, which is beneficial for organizations needing specific optimizations and stringent security protocols. In contrast, classic serverless databases like SQLite offer simplicity by eliminating the need for manual infrastructure management, making them ideal for rapid prototyping, embedded applications, or teams without dedicated DBAs. Neo-serverless databases strike a balance, offering managed services that eliminate the need for infrastructure management while providing some flexibility and scalability.
    \item \textbf{Scalability:} Server-based databases typically require manual intervention for scaling, which can be resource-intensive and costly, particularly during sudden increases in demand. Classic serverless databases like SQLite have limited scalability, as they are designed for lightweight, embedded use cases with modest scaling needs. Neo-serverless databases, on the other hand, are designed to scale automatically according to demand, making them ideal for cloud-based applications that experience fluctuating workloads without the need for manual configuration.
    \item \textbf{Cost Model:} Server-based databases generally involve substantial fixed costs related to infrastructure maintenance, irrespective of actual utilization. Classic serverless databases like SQLite are inherently cost-effective due to their minimal resource requirements and the absence of ongoing infrastructure expenses, making them suitable for applications with limited budgets or where infrastructure simplicity is paramount. Neo-serverless databases adopt a pay-as-you-go model, where costs are directly tied to usage, providing cost efficiency for applications with variable demands and reducing the overhead associated with maintaining dedicated infrastructure.
\end{itemize}

\subsubsection{Types of databases}

Broadly speaking, databases can be categorized into several types, with two of the most prominent being relational databases and document-based databases \cite{singh15TypesDatabases2024}. Relational databases, such as SQLite, use a schema-based, tabular structure to organize data into rows and columns, enforcing strict relationships through primary and foreign keys. This approach ensures consistency and reliability, making relational databases well-suited for applications requiring high levels of data integrity and well-defined relationships between data entities.

On the other hand, document-based databases, such as MongoDB, are designed to handle unstructured or semi-structured data. They store data in the form of documents, typically using Binary JSON (BSON), allowing for more flexible data models compared to relational databases. Document databases are particularly advantageous when dealing with data that may not fit neatly into predefined schemas, or when requirements are subject to change, offering greater agility in terms of data structure.

\subsubsection{SQLite: Relational Database}

SQLite is a lightweight, embedded relational database management system (RDBMS) that is self-contained and serverless. Unlike traditional server-based databases, SQLite operates directly within the application that accesses the database, making it a highly suitable option for small to medium-sized applications, rapid prototyping, mobile applications, and embedded systems. As a relational database, SQLite utilizes a structured schema to store data in tables consisting of rows and columns, akin to other RDBMSs such as MySQL or PostgreSQL. Importantly, SQLite fully supports ACID (Atomicity, Consistency, Isolation, Durability) transactions, ensuring data integrity and reliability even in the event of system failures or power outages \cite{SQLiteTransactional}. However, unlike server-based databases, SQLite is serverless, meaning it does not require a separate server process to manage transactions. Instead, SQLite directly interacts with ordinary disk files to read and write data, providing an efficient and simplified approach to data management without the overhead of server configuration. This serverless architecture, combined with ACID compliance, makes SQLite exceptionally easy to integrate and deploy, particularly for applications where simplicity, minimal configuration, data integrity, and low operational overhead are paramount \cite{SQLite}.

SQLite employs SQL (Structured Query Language) as its querying language \cite{SQLite}. SQL is a standardized language used to manage and manipulate relational databases, enabling operations such as data insertion, retrieval, updates, and deletions. SQLite's implementation of SQL is largely compliant with the SQL-92 standard, making it a familiar and accessible option for developers experienced with other relational databases. This compliance ensures that users can leverage their existing SQL knowledge to interact with SQLite effectively.

With regard to indexing, SQLite employs B-tree indexing to organize data and enhance query performance. B-tree indexes maintain data in a balanced tree structure, facilitating efficient searching, sorting, and range queries \cite{DatabaseFileFormat}. This indexing method ensures that data retrieval operations are optimized, reducing the need for costly full table scans, thereby significantly improving query performance. In addition to B-tree indexing, SQLite supports unique indexes to enforce data integrity by ensuring that specific columns contain unique values, and composite indexes to provide efficient access paths for queries involving multiple columns. These indexing capabilities are instrumental in maintaining optimal performance in SQLite databases, particularly when handling queries on larger datasets.

Below is an example of how to use SQLite with Python. This implementation demonstrates the creation of an SQLite database and inserting data into the database.

\vspace{0.1cm}

\noindent
\begin{minipage}{\linewidth}
\begin{lstlisting}[caption={Sample SQLite3 Code}]
import sqlite3

# Initialize the dataset
data = generate_data(num_rows)

# Connect with the database
conn = sqlite3.connect(db_name)

conn.execute('PRAGMA synchronous = OFF')     # Define optimizations
conn.execute('PRAGMA journal_mode = MEMORY') # Define optimizations

# Format the SQL insert statement
columns = ', '.join(f'col{i} INTEGER' for i in range(100))
cursor.execute(f'CREATE TABLE IF NOT EXISTS test_table ({columns})')
placeholders = ', '.join('?' for _ in range(100))
insert_query = f'INSERT INTO test_table VALUES ({placeholders})'

# Insert bulk amount of insert query
cursor.executemany(insert_query, data)

# Commit the transaction
conn.commit()

# Close the connection
conn.close()
\end{lstlisting}
\end{minipage}

\subsubsection{MongoDB: Document Database}

MongoDB is a NoSQL, document-oriented database management system specifically designed to facilitate the storage and retrieval of data in a flexible, non-relational format. As a member of the broader NoSQL database family, MongoDB employs a document-based model, utilizing BSON (Binary JSON) for data storage. This format supports dynamic and flexible data structures, enabling efficient handling of unstructured and semi-structured data. Unlike relational databases that organize data in tabular form with fixed schemas, MongoDB represents data as documents that resemble JSON objects, allowing for hierarchical relationships within a single data record. This document-centric architecture provides developers with significant flexibility, as fields within documents can vary across records. MongoDB's schema-less nature permits the addition of new fields without necessitating structural changes to the entire database, making it highly advantageous for applications where data models are subject to frequent evolution. Its scalability, distribution capabilities, and capacity to manage large volumes of diverse data make MongoDB particularly well-suited for modern web applications, big data initiatives, and environments with evolving data requirements \cite{MongoDBDeveloperData}.

The benefits of MongoDB in comparison to traditional databases are primarily rooted in its flexibility, scalability, and support for rapid development. The flexible document model of MongoDB allows for heterogeneous data structures without the need for predefined schemas, making it ideal for agile development environments where requirements are constantly changing. MongoDB is also designed for horizontal scalability, allowing it to accommodate growing data volumes by distributing data across multiple servers—a process known as sharding. This approach contrasts with traditional RDBMSs, which often require complex vertical scaling (enhancing the capacity of a single server). The distributed nature of MongoDB renders it particularly suitable for cloud-based deployments and big data projects that require seamless scalability. Additionally, MongoDB's document model enables developers to interact with data in a manner that closely mirrors the representation in the application layer, thereby simplifying data mapping and accelerating development.

MongoDB uses a proprietary query language based on a JSON-like syntax. This query language provides developers with the ability to execute a wide range of operations, including creating, reading, updating, and deleting documents. Its JSON-like syntax is both intuitive and easy to understand, particularly for developers who are familiar with JavaScript and web development, as it resembles working with native JSON objects. The query language includes advanced features such as aggregation pipelines, which support complex data transformations and analyses, and comprehensive indexing capabilities, enabling efficient data manipulation and retrieval.

In terms of indexing, MongoDB supports B-tree indexes to enhance the efficiency of data retrieval processes. Indexes in MongoDB are instrumental in optimizing query performance, and they can be created on any field within a document to facilitate rapid data access. In addition to standard B-tree indexes, MongoDB also offers a variety of specialized indexes, including compound indexes (which index multiple fields simultaneously), geospatial indexes (designed for location-based queries), text indexes (for full-text search capabilities), and hashed indexes (used to distribute data for sharding purposes). These diverse indexing options enable MongoDB to maintain high performance levels, even when managing large collections and complex queries, thereby enhancing the overall responsiveness of the database.

Below is an example of how to use MongoDB with Python. This implementation demonstrates how to connect to a MongoDB server and insert data into the database.

\vspace{0.1cm}

\noindent
\begin{minipage}{\linewidth}
\begin{lstlisting}
from pymongo import MongoClient

# Initialize the dataset
data = generate_data(num_rows)

# Connect with the database
client = MongoClient('mongodb://localhost:27017/')

# Select the database and the collection
db = client[db_name]
collection = db['test_collection']

# Insert the data
collection.insert_many(data)
\end{lstlisting}
\end{minipage}

\subsection{Limitations of Traditional Data Storage Formats}

In previous sections, we introduced traditional data storage formats, including CSV, JSON, and TXT, as well as database management systems like SQLite and MongoDB. Each of these formats exhibits inherent strengths and limitations, shaped largely by their design and intended use cases. In the following sections, we will discuss the specific challenges associated with these formats, with a focus on their implications for efficiency, performance, and usability, particularly in complex, large-scale data environments.

One significant limitation is the serialization/deserialization overhead. Serialization is the process of converting data into a format suitable for storage, while deserialization reconstructs the data into a usable form for applications. Many formats, including JSON, CSV, TXT, and binary formats, support serialization, but traditional text-based formats are not optimized for efficient deserialization, resulting in significant performance bottlenecks. This inefficiency leads to computationally intensive operations, especially for large datasets, thereby increasing latency and reducing overall performance. Similarly, CSV and TXT formats, while simple, lack efficient mechanisms for parsing and converting data, particularly when dealing with nested or hierarchical structures, further exacerbating performance issues. As a result, storage formats that allow for different encoding methods are preferred for performance optimization. For instance, SQLite and MongoDB allow for basic encodings for primitive data types such as bool, ints, floats, and text.

Another major limitation, closely related to serialization and deserialization challenges, involves the handling of complex data types. Traditional formats like CSV are well-suited for storing simple, tabular data but struggle to accommodate more intricate data structures. While JSON provides more flexibility, it is not particularly efficient at managing deeply nested or highly complex objects, leading to increased difficulties in data management. Relational databases such as SQLite, on the other hand, enforce rigid schema definitions, which may not easily accommodate evolving or dynamic data structures, making them less suitable for modern applications where data requirements frequently change. Some solutions, such as using text fields or Binary large obects (BLOB) fields, can address these limitations to some degree, but they introduce their own challenges. Text fields add deserialization overhead, while BLOB fields make the data application-dependent, reducing portability, as other users require the same application to decode the BLOB.

A further  limitation is the dependency on indexing for performance. Both traditional databases and data storage formats like SQLite and MongoDB rely heavily on indexing to improve query efficiency. Although indexing can significantly accelerate data retrieval, creating and maintaining indexes introduces overhead, particularly in write-heavy applications. This dependence on indexing also highlights a fundamental inefficiency: in scenarios where indexes are absent or improperly optimized, query performance can degrade substantially, affecting overall system responsiveness.

Portability also presents a significant challenge for traditional data storage formats. While formats such as CSV and JSON are inherently portable, server-based databases like MongoDB often entail dependencies that complicate the process of transferring or replicating data across different systems. The absence of a standardized migration mechanism between distinct environments further exacerbates the difficulty of ensuring data portability, particularly when managing large datasets.

Lastly, the complex setup and implementation associated with traditional database management systems can pose considerable barriers. Relational databases typically require extensive setup, configuration, and ongoing maintenance, making them unsuitable for small-scale or rapid development projects. As seen in the case of SQLite, there is substantial overhead involved in crafting suitable queries, and despite SQLite's simplicity compared to other relational databases, its setup can still be cumbersome compared to more modern, out-of-the-box solutions. In the case of MongoDB, while the coding implementation is relatively straightforward, maintaining the necessary server infrastructure or relying on cloud services adds an additional layer of complexity to the deployment process.

These limitations underscore the need for a more efficient, scalable, and portable data storage solution. The next section will explore how the Parquet format and the ParquetDB implementation address these challenges, providing a robust and user-friendly alternative to traditional data storage methods.

\section{ParquetDB}

ParquetDB is a lightweight, Python-based data management system that builds upon Apache Parquet files \cite{Parquet} utilizing PyArrow \cite{PythonApacheArrow}. It offers an advanced and efficient approach for managing complex data types, utlizing the benefits of columnar storage to improve data compression and retrieval efficiency. ParquetDB overcomes the limitations mentioned in the previous section by minimizing serialization overhead, thus addressing the inefficiencies typically found in traditional data storage formats. By mitigating these bottlenecks, ParquetDB provides a robust mechanism for managing large-scale datasets where rapid access and manipulation are essential.



In this section, we will explore the following topics:

\begin{itemize}[label={--},leftmargin=5pt, labelsep=3pt]
    \item \textbf{The Apache Parquet Format:}  A comprehensive overview of the Parquet file format, its features, and its suitability for managing large datasets
    \item \textbf{The Design and Architecture of ParquetDB:} An in-depth look at how ParquetDB is structured, including its core components and how it leverages Parquet and PyArrow to optimize data storage and retrieval.
    \item \textbf{Examples of Using ParquetDB:} Detailed examples demonstrating the effective use of ParquetDB for textbf data management, highlighting its core functionalities and diverse applications.
    \item \code{Other Important Notes:} Additional considerations, including best practices, limitations, and recommendations for effectively integrating ParquetDB into data workflows.
\end{itemize}

\subsection{Apache Parquet Format}

At the heart of ParquetDB's architecture is the use of Apache Parquet as the storage format. Parquet is a highly efficient columnar storage format specifically designed to address the complexities of fast query performance and efficient deserialization, particularly for large-scale, complex datasets. This layer directly addresses common data-handling challenges like serialization overhead, allowing ParquetDB to store and query complex datasets in a highly efficient manner. The benefits of parquet files are:

\begin{itemize}[label={--},leftmargin=5pt, labelsep=3pt]
    \item \textbf{Columnar Format:} Parquet organizes data by columns instead of rows, making it ideal for analytical queries that require only subsets of columns. This format allows for better compression and reduces I/O overhead.
    
    \item \textbf{Schema Embedding:} Each Parquet file contains a schema that describes the data, ensuring consistency and facilitating schema evolution.
    
    \item \textbf{Predicate Pushdown:} Parquet's structure allows ParquetDB to optimize queries by reading only relevant data blocks.

    \item \textbf{Predicate Pushdown:} Parquet's structure allows to only read the relevant columns into memory.
    
    \item \textbf{Efficient Encoding and compression:} Parquet files allow for column-level encoding and compression techniques improving both read performance and storage efficiency.
    \item \textbf{Metadata support:} Parquet files are able store store table- and column- level metadata.
    
    \item \textbf{Batching support:} Parquet Files handles hdata in groups of column data making them conducive to batching operations
\end{itemize}

By using Parquet files, ParquetDB optimizes the serialization and deserialization process, providing fast and scalable data access that fits seamlessly into machine learning and big data pipelines.

\begin{figure}[t!]
    \centering
    \includegraphics[width=0.8\textwidth]{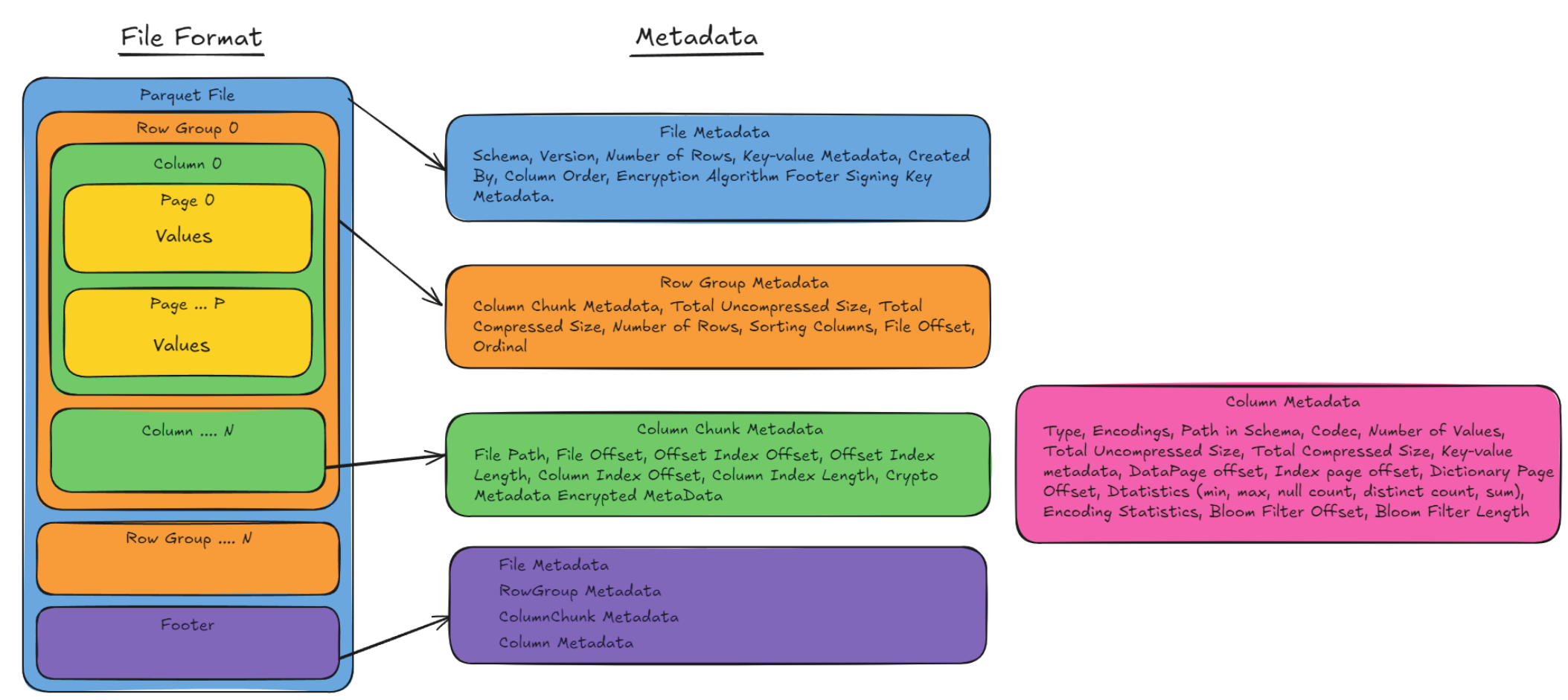}
    \caption{Parquet File Format Overview. This diagram illustrates the structure of a Parquet file, including Row Groups, Columns, Pages, and the Footer. The metadata associated with each level provides essential details, such as schema, offsets, compression sizes, encryption, and statistical summaries. These metadata components enable efficient data storage, retrieval, and filtering, making Parquet an ideal choice for analytics and big data processing.}
    \label{fig:parquet-file-format}
\end{figure}

The structure of a Parquet file is depicted in Figure \ref{fig:parquet-file-format}, where files are partitioned into row groups, which are further divided into column chunks. Each column chunk is composed of pages, representing the smallest unit of storage. A footer at the end of each file consolidates critical metadata, including schema information and statistics for each row group, such as minimum and maximum values. This hierarchical organization of row groups, column chunks, and pages supports advanced features such as predicate pushdown and columnar projection, while facilitating efficient data compression and retrieval.

\begin{figure}[t]
    \centering
    \includegraphics[width=0.8\textwidth]{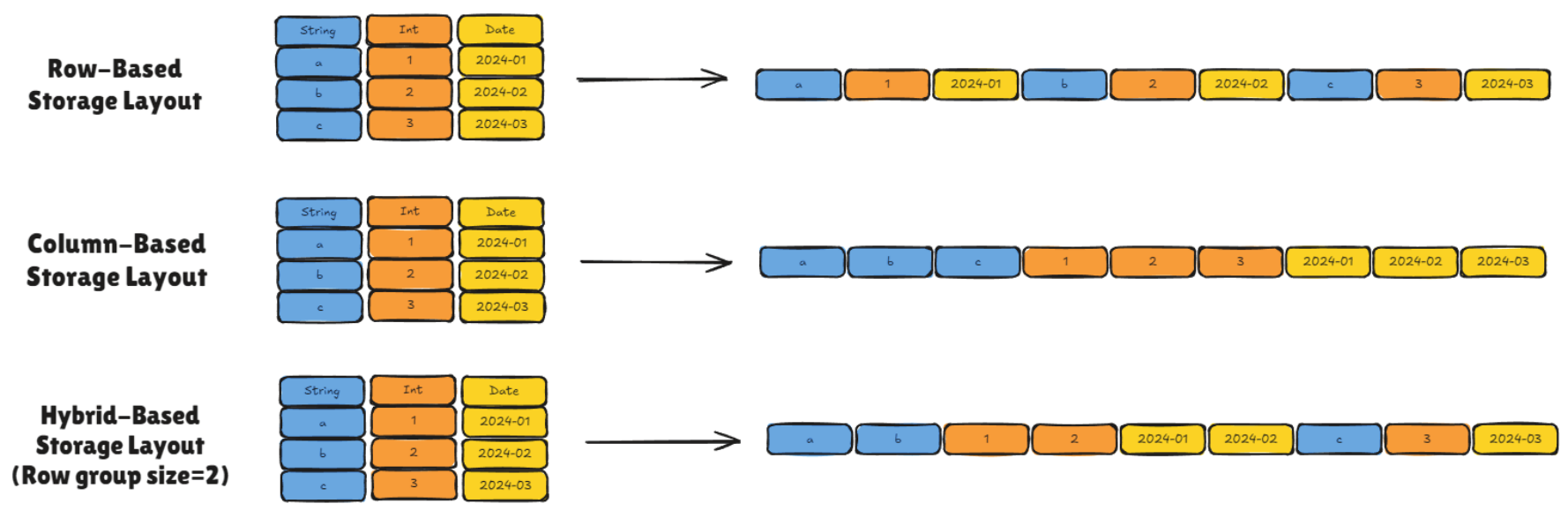}
    \caption{Comparison of Storage Layouts. Row-Based, Column-Based, and Hybrid-Based (Row Group Size = 2). Parquet files utilize a hybrid storage layout, balancing the strengths of row-based and column-based storage by grouping rows together for efficient read and write operations, making them suitable for analytic queries on large datasets.}
    \label{fig:parquet-storage-layout}
\end{figure}

Traditional data storage formats often involve a compromise between row-based storage, optimized for transactional workloads, and column-based storage, which excels in analytical queries. Parquet introduces a hybrid storage model that leverages the benefits of both paradigms, enabling data to be stored in a columnar format while maintaining the efficiency required for diverse query types. To fully appreciate the advantages of Parquet's hybrid approach, it is crucial to understand the distinction between traditional row-based and column-based formats (see Figure \ref{fig:parquet-storage-layout} for an illustration of data organization and access differences in these formats). Row-based formats, such as CSV or relational databases, store data sequentially by row, making them ideal for workloads requiring frequent row-level operations like updates or inserts. However, they are inefficient for analytical queries, which necessitate scanning entire rows even when only specific columns are required. In contrast, column-based formats, commonly used in data warehouses, store values from the same column together, enabling faster aggregations and more efficient analytics. Parquet's predominantly columnar approach offers a more efficient means of data retrieval, effectively bridging the gap between transactional and analytical workloads.

Parquet files utilize metadata extensively to optimize data access. This metadata includes schema information, statistics, and offsets, which enable query engines to make informed decisions about data retrieval. For instance, detailed statistics such as minimum and maximum values for each column facilitate predicate pushdown, allowing query engines to bypass irrelevant row groups or pages based on filter conditions. Schema metadata ensures data type consistency, mitigating serialization errors and enhancing interoperability across different platforms. This combination of rich metadata makes Parquet files self-descriptive, which significantly improves both usability and performance.

Through detailed metadata, Parquet optimizes data retrieval using several key mechanisms. By storing statistics for row groups, Parquet enables efficient skipping of irrelevant data blocks, thus facilitating effective predicate and projection pushdown. Predicate pushdown allows the query engine to filter data directly at the storage level, skipping over data blocks that do not meet the filter criteria, similar to applying a WHERE clause in SQL. Projection pushdown, in contrast, ensures that only the required columns are retrieved, akin to using the SELECT statement in SQL. By minimizing the volume of data processed and loaded into memory, both predicate and projection pushdown significantly enhance query performance and reduce computational overhead, making Parquet an ideal choice for big data applications.

Building upon the concept of metadata optimization, Parquet achieves high efficiency by minimizing data size through encoding and compression techniques. Encoding reduces data size while preserving direct access to each data point, whereas compression further reduces data footprint, albeit requiring decompression before use. Parquet supports a range of encoding techniques, each optimized for specific data types, including Plain Encoding, Dictionary Encoding, Run-Length Encoding/Bit-Packing Hybrid, Delta Encoding, Delta-Length Byte Array, Delta Strings, and Byte Stream Split. For compression, Parquet offers several options such as GZIP, LZO, BROTLI, ZSTD, and LZ4\_RAW. These methods work complementarily to significantly decrease data size, thereby saving storage space and optimizing performance for data transfer operations.

Apache Parquet's robust architecture, rich metadata, and sophisticated encoding and compression strategies make it an ideal choice to be used as a basis of a database. As we move on to the next section, we will explore the design and architecture of ParquetDB. This examination will provide more understanding of how Parquet's strengths are used to create an efficient, portable database solution. For more information on the Parquet File format please refer to the Supplemental Information (SI).

\subsection{Integration of PyArrow in ParquetDB}

PyArrow serves as the computational backbone of ParquetDB, providing fast in-memory processing of data stored in Parquet format while including utilities for handling complex data types, performing efficient filtering and casting operations, and parallelizing I/O tasks. It enables efficient in-memory data manipulation, crucial for machine learning and data-intensive workflows, and its high-performance reading and writing capabilities minimize the overhead of frequent data operations. Since the core of PyArrow is written in C++, it allows for parallelism on threads by ignoring the Global Interpreter Lock (GIL) of Python, which often restricts pythons performances for parallel tasks. Additionally, PyArrow's API compatibility with tools like Apache Spark and Dask creates integration of ParquetDB into existing big data pipelines.

\newpage

\subsection{ParquetDB interface}

{
\setlength{\intextsep}{0pt}
\begin{table}[H]
    \caption{Arguments and descriptions for the initialization of the \texttt{ParquetDB} class. Example code is listed below the table.}
    \begin{adjustbox}{width=\textwidth,center}
    \centering
    \begin{tblr}{
        colspec = {X[0.30\textwidth,c,m]|X[0.70\textwidth,l,m]},
        column{1} = {valign=m, halign=c}, 
        column{2} = {halign=l},
        cell{2}{2} = {halign=c}, 
        cells = {font=\footnotesize\selectfont},
        stretch = 0,
        rowsep = 3pt,
        hlines = {black, 1pt},
        vlines = {black, 1pt},
    }
    \SetCell[c=2]{c}\textbf{ParquetDB} \\ 
    \textbf{Argument} & \textbf{Description} \\ 
    \textbf{db\_path} & The path to the dataset to be created or accessed. \\ 
    \textbf{initial\_fields} & A list of fields to be added to the dataset. \\ 
    \textbf{serialize\_python\_objects} & If \code{True}, the dataset will be serialized using the \code{pickle} module. \\ 
    \textbf{use\_multiprocessing} & If \code{True}, the dataset will be processed in parallel using the \code{multiprocessing} module. \\ 
    \end{tblr}
    \end{adjustbox}

    \centering

    \begin{minipage}{\textwidth}

    \begin{lstlisting}
# Example usage:
db = ParquetDB(db_path='/path/to/db')
    \end{lstlisting}
    \end{minipage}

    \label{table:parquetdb-init}
\end{table}
}
The \code{ParquetDB} class serves as the main interface to interact with the database. It provides methods for CRUD (Create, Read, Update, Delete) operations while abstracting away the complexities of working directly with Parquet files. It also provide other useful convience methods for interacting with the dataset. The main methods to interact with the database are:

\begin{itemize}[label={--},leftmargin=5pt, labelsep=3pt]
    \item \textbf{Create:} The \code{create} method accepts data in various formats (pandas DataFrame, dict, list) and writes the data to a new Parquet file. It also handles schema alignment and data normalization, ensuring new data is integrated into the existing dataset. More information about the \code{create} method can be found in Table \ref{table:create-function}.
    {
    \setlength{\intextsep}{0pt}
    \begin{table}[H]
        \caption{Arguments and descriptions for the \code{create} function. Example code is listed below the table.}
        \begin{adjustbox}{width=\textwidth,center}
        \centering
        \begin{tblr}{
            colspec = {X[0.15\textwidth,c,m]|X[0.75\textwidth,l,m]},
            column{1} = {valign=m, halign=c}, 
            column{2} = {halign=l}, 
            cell{2}{2} = {halign=c}, 
            cells = {font=\scriptsize\linespread{0.8}\selectfont},
            stretch = 0,
            rowsep = 3pt,
            hlines = {black, 1pt},
            vlines = {black, 1pt},
          }
        \SetCell[c=2]{c}\textbf{ParquetDB.create} \\ 
        \textbf{Argument} & \textbf{Description} \\ 
        \textbf{data} & The data to be added to the database. Can be a dict, a list of dicts, \code{pandas.DataFrame}. \\
        \textbf{schema} & The schema for the incoming data. Accepts a \code{pyarrow.Schema} object (optional). By default, the schema will be infered during the input process, but a schema can be provided to enforce a schema on the input data. The schema has to be compataible with the input data. \\
        \textbf{metadata} & Metadata to be attached to the table. Must be a dictionary (optional). \\
        \textbf{fields\_metadata} & Metadata to be attached to the fields. Must be a dictionary (optional). \\
        \textbf{normalize\_ dataset} & If True, the dataset will be normalized after being added. Default is False (optional). \\
        \textbf{normalize\_ config} & Configuration for normalization, optimizing performance by managing row distribution and file structure. Accepts a \code{NormalizeConfig} object (optional). \\
        \textbf{treat\_fields\_as\_ragged} & A list of fields to treat as ragged arrays. Default is \code{None}. \\
        \textbf{convert\_to\_fixed\_shape} & If True, the ragged arrays will be converted to fixed shape arrays. Default is True. \\
        \end{tblr}
        \end{adjustbox}

        \centering
        \begin{minipage}{\textwidth}
        \begin{lstlisting}
    # Example 
    my_data = [{'name': 'John', 'age': 30}, {'name': 'Jane', 'age': 25}]
    my_schema = pa.schema([('name', pa.string()), ('age',pa.int64())])
    
    db.create(data=my_data, schema=my_schema, metadata={'source': 'api'})\end{lstlisting} 
        \end{minipage}
    
        \label{table:create-function}
    \end{table}
    \setlength{\intextsep}{10pt}
    }

\newpage
    \item \textbf{Read:} The \code{read} method retrieves data based on filters, column selection, and format (either as a full PyArrow Table, PyArrow Record Batch, or PyArrow Dataset). This method supports both column and predicate pushdown to optimize retrieval. Additional, loading options can be provided to control the batch size, the batch readahead, the fragment readahead, and the use of threads. More information about the \code{read} method can be found in Table \ref{table:read-function}.

{
\setlength{\intextsep}{0pt}
\begin{table}[H]
    \caption{Arguments and descriptions for the \code{read} function. Example code is listed below the table.}
    \begin{adjustbox}{width=\textwidth,center}
    \centering
    \begin{tblr}{
        colspec = {X[0.15\textwidth,c,m]|X[0.75\textwidth,l,m]},
        column{1} = {valign=m, halign=c}, 
        column{2} = {halign=l},
        cell{2}{2} = {halign=c}, 
        cells = {font=\scriptsize\linespread{0.8}\selectfont},
        stretch = 0,
        rowsep = 3pt,
        hlines = {black, 1pt},
        vlines = {black, 1pt},
    }
    \SetCell[c=2]{c}\textbf{ParquetDB.read} \\ 
    \textbf{Argument} & \textbf{Description} \\ 
    \textbf{load\_format} & Determines whether the data is returned as a full PyArrow Table, batches, or a PyArrow Dataset. For \code{batches}, returns a generator. For \code{table}, returns a full PyArrow Table. For \code{dataset}, returns a PyArrow Dataset. \\
    \textbf{batch\_size} & Allows control over how much data is processed per batch. \\ 
    \textbf{ids} & A list of specific record IDs to retrieve. \\ 
    \textbf{columns} & Specify certain columns to include or exclude. \\ 
    \textbf{include\_cols} & Controls whether the specified columns are included or excluded. \\ 
    \textbf{filters} & Allows users to apply a list of PyArrow compute expressions for conditional data retrieval. The expressions provided will be combined together with the bitwise 'and' operator. \\
    \textbf{rebuild\_ nested\_ struct} & If \code{True}, rebuilds the nested structure (default: \code{False}). \\ 
    \textbf{rebuild\_ nested\_ from\_ scratch} & If \code{True}, rebuilds the nested structure from scratch (default: \code{False}). \\ 
    \textbf{load\_config} & Configuration for loading data, optimizing performance by managing memory usage. \\ 
    \textbf{normalize\_ config} & Configuration for the normalization process, optimizing performance by managing row distribution and file structure. \\    
    \end{tblr}
    \end{adjustbox}

    \centering
    \begin{minipage}{\textwidth}
    \begin{lstlisting}
# Example usage:
load_config = LoadConfig(batch_size=100, batch_readahead=8, fragment_readahead=2)
result = db.read(load_format='batches', load_config=load_config,
                 columns=['name', 'age'], filters=[('age', '>=', 25)])
for batch in result:
    process(batch)\end{lstlisting}
    \end{minipage}

    \label{table:read-function}
\end{table}
}

    \item \textbf{Update:} The \code{update} method modifies existing records in the database. It leverages schema casting and normalization to ensure that the updated data remains consistent with the rest of the dataset. More information about the \code{update} method can be found in Table \ref{table:update-function}.
    
{
\setlength{\intextsep}{0pt}
\begin{table}[H]
    \caption{Arguments and descriptions for the \code{update} function. Example code is listed below the table. \code{update} is used to update existing records in the database. Unlike the \code{create} method, the input \textbf{data} must contain \code{id} for the updates to be applied.}
    \begin{adjustbox}{width=\textwidth,center}
    \centering
    \begin{tblr}{
        colspec = {X[0.15\textwidth,c,m]|X[0.75\textwidth,l,m]},
        column{1} = {valign=m, halign=c}, 
        column{2} = {halign=l}, 
        cell{2}{2} = {halign=c}, 
        cells = {font=\scriptsize\linespread{0.8}\selectfont},
        stretch = 0,
        rowsep = 3pt,
        hlines = {black, 1pt},
        vlines = {black, 1pt},
    }
    \SetCell[c=2]{c}\textbf{ParquetDB.update} \\ 
    \textbf{Argument} & \textbf{Description} \\ 
    \textbf{data} & Can be a \code{dict}, a \code{list of dicts}, or a \code{pandas.DataFrame} \\ 
    \textbf{schema} & By default, the schema will be inferred during the input process, but a schema can be provided to enforce compatibility with the input data. \\ 
    \textbf{metadata} & Allows attaching metadata to the table if needed. \\ 
    \textbf{fields\_metadata} & Allows attaching metadata to the fields if needed. \\
    \textbf{update\_keys} & The keys to use for the update. If a list, the update will be performed on the intersection of the existing data and the incoming data. \\
    \textbf{treat\_fields\_as\_ragged} & A list of fields to treat as ragged arrays. \\
    \textbf{convert\_to\_fixed\_shape} & If True, the ragged arrays will be converted to fixed shape arrays. \\
    \textbf{normalize\_ config} & Configuration for the normalization process, optimizing performance by managing row distribution and file structure. Accepts a \code{NormalizeConfig} object (optional). \\
    \end{tblr}
    \end{adjustbox}

    \centering

    \begin{minipage}{\textwidth}

    \begin{lstlisting}
# Example usage:
data = { 'name': ['Alice', 'Bob', 'Charlie'], 'age': [25, 30, 35]}

db.update(data=data,  metadata={'source': 'manual'}, 
          normalizeConfig=NormalizeConfig())\end{lstlisting}
    \end{minipage}

    \label{table:update-function}
\end{table}
}

\newpage
    \item \textbf{Delete:} The \code{delete} method removes records or columns from the dataset. It enforces constraints such as preventing the deletion of primary keys or essential columns. More information about the \code{delete} method can be found in Table \ref{table:delete-function}.
{
\setlength{\intextsep}{0pt}
    \begin{table}[H]
        \caption{Arguments and descriptions for the \code{delete} function. Example code is listed below the table.}
        \begin{adjustbox}{width=\textwidth,center}  
        \centering
        \begin{tblr}{
            colspec = {X[0.15\textwidth,c,m]|X[0.75\textwidth,l,m]},
            column{1} = {valign=m, halign=c}, 
            column{2} = {halign=l}, 
            cell{2}{2} = {halign=c}, 
            cells = {font=\scriptsize\linespread{0.8}\selectfont},
            stretch = 0,
            rowsep = 3pt,
            hlines = {black, 1pt},
            vlines = {black, 1pt},
        }
        \SetCell[c=2]{c}\textbf{ParquetDB.delete} \\ 
        \textbf{Argument} & \textbf{Description} \\ 
        \textbf{ids} & A list of record IDs to delete from the database. \\ 
        \textbf{columns} & A list of column names to delete from the dataset (optional). If not provided, it will be inferred from the existing data. Default: \code{None}. \\ 
        \textbf{filters} & Allows users to apply a list of PyArrow compute expressions for conditional data removal. \\
        \textbf{normalize\_ config} & Configuration for the normalization process, optimizing performance by managing row distribution and file structure. Accepts a \code{NormalizeConfig} object (optional). Default: \code{NormalizeConfig()}. \\
        \end{tblr}
        \end{adjustbox}
    
        \centering
        \begin{minipage}{\textwidth}
    
        \begin{lstlisting}
    # Example usage:
    from parquetdb import NormalizeConfig
    
    # Delete specific records by ID
    db.delete(ids=[1, 2, 3])
    
    # Delete specific columns from the dataset
    db.delete(columns=['age', 'salary'])
    
    # Delete records and normalize
    normalize_config = NormalizeConfig(max_rows_per_file=500)
    db.delete(ids=[4, 5, 6], normalize_config=normalize_config)\end{lstlisting}
        \end{minipage}
    
        \label{table:delete-function}
    \end{table}   
}

\end{itemize}

\subsection{Data flow in ParquetDB}

Following the discussion of the core methods in the API, we now explore how ParquetDB manages data flow. The data flow in ParquetDB is designed to handle various data formats and convert them into a unified format for efficient processing and storage. Whether the input data is provided as a Python list of dictionaries, a dictionary of column arrays, a pandas DataFrame, or a PyArrow Table, ParquetDB standardizes the data into a PyArrow Table before any further operations are performed. This conversion ensures consistency and allows ParquetDB to use PyArrow's data handling capabilities. Below is an outline of the typical data flow when using the create or update functions in ParquetDB:

\subsubsection{Input Data Formats}
ParquetDB is flexible in terms of the input formats it accepts. The input to the create and update functions can be in any of the following formats:
\begin{itemize}[label={--},leftmargin=5pt, labelsep=3pt]
    \item \textbf{Python List of Dictionaries:} A list where each element is a dictionary representing a record (row) with key-value pairs corresponding to column names and data.
    \item \textbf{Python Dictionary of Column Arrays:} A dictionary where the keys represent column names, and the values are arrays (lists) of data, with each array representing the data for a column.
    \item \textbf{Pandas DataFrame:} A popular tabular data structure used in data science, representing the data as rows and columns, which ParquetDB can convert into a PyArrow Table.
    \item \textbf{PyArrow Table:} A native PyArrow format where the data is already structured for in-memory processing, offering high efficiency when working with ParquetDB.
\end{itemize}
Regardless of the input format, the first step in the data flow is the conversion of the input data into a PyArrow Table.

\subsubsection{Preprocessing Incoming data}

Once the input data is in the form of a PyArrow Table, ParquetDB performs several preprocessing steps to ensure the data is correctly formatted and compatible with the Parquet format:

\begin{itemize}[label={--},leftmargin=5pt, labelsep=3pt]
    \item \textbf{Handling Empty Structs:}
        In Parquet, empty nested structures such as dictionaries or structs cannot be stored directly, posing a limitation for data integrity. To address this, ParquetDB introduces a dummy variable into any empty struct fields, ensuring that all fields, including those that are nested and potentially empty, contain valid data. This approach is essential for preserving the integrity of the dataset during storage and maintaining compatibility with the Parquet file format.

    \item \textbf{Flattening Nested Structures:}
        ParquetDB simplifies nested data structures, such as structs or dictionaries with nested fields, by flattening them into a flat table, making the data model easier to process and query. During this process, new columns are generated for each nested field, with names that reflect the original structure using a naming convention like \code{parent.child1.child2} for deeply nested fields. For example, a field named \code{address} with nested fields \code{city} and \code{postal\_code} would result in columns named \code{address.city} and \code{address.postal\_code}, streamlining data access and analysis. Flattening these structures simplifies the schema but also makes it easier to handle changes and updates in the data. Working with a flat structure allows ParquetDB to detect changes in the data more effectively and ensures better compatibility with query engines and data processing pipelines.

    \item \textbf{Column Reordering:}
        ParquetDB alphabetically orders columns to simplify schema change detection and updates. New columns introduced through updates or changes are automatically placed in their correct position, making differences between schemas easier to identify.

    \item \textbf{Schema Alignment:} 
        The incoming data is compared against the existing schema of the dataset. If there are schema differences, ParquetDB either updates the schema or casts the data to fit the existing schema, depending on the operation.
    
    \item \textbf{Writing to Parquet:} 
        Finally, the data is written to disk as a Parquet file, using PyArrow's efficient file-writing utilities. At this stage, ParquetDB may also invoke the normalization process, ensuring that the data is evenly distributed across files and that any schema changes, updates, or deletions are reflected consistently across the entire dataset.
\end{itemize}

\subsection{Create, Update, Delete Processing}

\subsubsection{Create Processing}

When data is first input into ParquetDB, it undergoes several steps:

\begin{enumerate}
    \item \textbf{Pre-processing}: The data is prepared according to the pre-processing steps defined in ParquetDB.

    \item \textbf{ID Generation}:
    \begin{itemize}
        \item ParquetDB generates unique IDs for each new record.
        \item If it's the first data input, IDs start at 0.
        \item For subsequent inputs, IDs continue from the current maximum ID in the database.
        \item These IDs are appended to the incoming data table in the \code{id} column
    \end{itemize}

    \item \textbf{Storage Process}:
    \begin{itemize}
        \item ParquetDB checks if the dataset directory is empty.
        \item If empty:
        \begin{itemize}
            \item A new Parquet file is created with the name \newline\code{\{dataset\_name\} \_0.parquet}.
        \end{itemize}
        \item If not empty:
        \begin{itemize}
            \item A schema reconciliation process takes place.
            \item If there's no schema change:
            \begin{itemize}
                \item The incoming data is stored in a new Parquet file named \code{\{dataset\_name\}\_\{i\}.parquet}, where \code{i} is the next available index.
            \end{itemize}
            \item If there's a schema change:
            \begin{itemize}
                \item Existing data is rewritten to align with the new schema.
                \item Incoming data is still stored in a new Parquet file as above.
            \end{itemize}
        \end{itemize}
    \end{itemize}

    \item \textbf{Normalization} (Optional):
    \begin{itemize}
        \item If \code{normalize\_dataset} is set to \code{True}:
        \begin{itemize}
            \item All Parquet files are rewritten to normalize the row distribution across files.
            \item This optimization can improve query performance.
        \end{itemize}
        \item By default, \code{normalize\_dataset} is \code{False} to reduce creation time.
        \item It's generally recommended to normalize the dataset after all create operations are complete.
    \end{itemize}

    \item \textbf{Backup Creation}:
    \begin{itemize}
        \item Before any modifications, current files are copied to a temporary directory.
        \item This backup ensures data recovery in case of errors during the create process.
    \end{itemize}

    \item \textbf{Error Handling}:
    \begin{itemize}
        \item If an error occurs during the modifications:
        \begin{itemize}
            \item The process is halted.
            \item Original files are restored from the temporary backup.
        \end{itemize}
    \end{itemize}
    
\end{enumerate}

This process ensures efficient data storage and maintains schema consistency across the dataset. The optional normalization step can be used to optimize performance, especially for larger datasets or frequent query operations.

\subsubsection{Update Process}

The update function in ParquetDB follows a careful process to modify existing data while maintaining data integrity and allowing for schema changes. Here's a step-by-step breakdown of the process:

\begin{enumerate}
    \item \textbf{Pre-processing}:
    \begin{itemize}
        \item Incoming data undergoes the pre-processing steps defined earlier in ParquetDB.
    \end{itemize}

    \item \textbf{Update Execution}:
    \begin{itemize}
        \item The system iterates through the columns in the incoming data table.
        \item It detects any changes in schema or data values.
    \end{itemize}

    \item \textbf{Schema Handling}:
    \begin{itemize}
        \item If new fields are present in the incoming data:
        \begin{itemize}
            \item These fields are added to the existing schema.
            \item Null values are inserted for existing records not present in the incoming data.
        \end{itemize}
    \end{itemize}

    \item \textbf{Data Modification}:
    \begin{itemize}
        \item The dataset is updated with the incoming data based on matching IDs.
    \end{itemize}

    \item \textbf{Normalization} (Optional):
    \begin{itemize}
        \item The update process utilizes the normalization function.
        \item Users can provide arguments to optimize the dataset during this process.
    \end{itemize}

    \item \textbf{Backup Creation}:
    \begin{itemize}
        \item Before any modifications, current files are copied to a temporary directory.
        \item This backup ensures data recovery in case of errors during the update process.
    \end{itemize}
    
    \item \textbf{Error Handling}:
    \begin{itemize}
        \item If an error occurs during the update:
        \begin{itemize}
            \item The process is halted.
            \item Original files are restored from the temporary backup.
        \end{itemize}
    \end{itemize}
\end{enumerate}

This process ensures that updates are applied safely and efficiently, allowing for both data modifications and schema evolution. The optional normalization step provides an opportunity to maintain or improve dataset performance during updates.

\subsubsection{Delete Processing}

The delete function in ParquetDB offers two distinct types of deletions: row-based (by \code{id}) and column-based. Here's an overview of the process:

\begin{enumerate}
    \item \textbf{Delete Type Detection}:
    \begin{itemize}
        \item ParquetDB first determines whether the deletion is:
        \begin{itemize}
            \item Row-based (\texttt{id} deletion)
            \item Column-based (\texttt{column} deletion)
        \end{itemize}
    \end{itemize}
    
    \item \textbf{Exclusive Operation}:
    \begin{itemize}
        \item Row and column deletions are mutually exclusive operations.
        \item They cannot be performed simultaneously in a single delete operation.
    \end{itemize}
    
    \item \textbf{Deletion Execution}:
    \begin{itemize}
        \item Based on the detected type, ParquetDB proceeds with either:
        \begin{itemize}
            \item Removing specified rows (by \texttt{id})
            \item Eliminating designated columns
        \end{itemize}
    \end{itemize}
    
    \item \textbf{Normalization} (Optional):
    \begin{itemize}
        \item The delete process utilizes the normalization function.
        \item Users can provide arguments to optimize the dataset during this process.
        \item This step can help maintain or improve performance after deletions.
    \end{itemize}

    \item \textbf{Backup Creation}:
    \begin{itemize}
        \item Before any modifications, current files are copied to a temporary directory.
        \item This backup ensures data recovery in case of errors during the update process.
    \end{itemize}
    
    \item \textbf{Error Handling}:
    \begin{itemize}
        \item If an error occurs during the update:
        \begin{itemize}
            \item The process is halted.
            \item Original files are restored from the temporary backup.
        \end{itemize}
    \end{itemize}
    
\end{enumerate}

This process ensures that deletions are carried out efficiently while maintaining the integrity of the database structure. The optional normalization step allows for dataset optimization, which can be particularly useful after large-scale deletions.


\newpage

\subsection{Advanced Topics and Features}

In this section we will talk about some advanced topics and featurs in ParqeutDB

\subsubsection{Rebuilding Nested Structures}

As discussed in the Data Flow section, input data is flattened for storage. While this approach is efficient, it may be disadvantageous in certain scenarios where users prefer to retain the original nested structure. To address this, ParquetDB provides methods to access data in its nested form when the \code{read} method is invoked. By specifying \code{rebuild\_nested\_struct=True}, users can reconstruct the nested structure in a separate directory named \code{\{dataset\_name\}\_nested}. Once the nested structure is rebuilt, ParquetDB will use this dataset for querying, allowing users to work with data in its original nested form. However, if any modifications are made after the initial call to \code{rebuild\_nested\_struct=True}, the nested structure must be regenerated by setting \code{rebuild\_nested\_from\_scratch=True}. This process will delete the existing nested directory, enabling ParquetDB to rebuild it with the latest modifications.

\subsubsection{LoadConfig class}

In the previous section, you may have noticed the keyword argument \code{load\_config}. This class is used to configure the loading process. It allows for control over the batch size, the batch readahead, the fragment readahead, and the use of threads. This can be useful to reduce RAM usage. More information about the \code{LoadConfig} class can be found in Table \ref{table:load-config}.

\begin{table}[hbt]
    \caption{Arguments and descriptions for the \code{LoadConfig} class. This class provides configuration options for loading data efficiently by specifying columns, filters, batch size, and memory usage.}
    \begin{adjustbox}{width=\textwidth,center}
    \centering
    \begin{tblr}{
        colspec = {X[0.15\textwidth,c,m]|X[0.75\textwidth,l,m]},
        column{1} = {valign=m, halign=c}, 
        column{2} = {valign=m, halign=l},
        cell{2}{2} = {halign=c}, 
        cells = {font=\scriptsize\linespread{0.8}\selectfont},
        stretch = 0,
        rowsep = 3pt,
        hlines = {black, 1pt},
        vlines = {black, 1pt},
    }
    \SetCell[c=2]{c}\textbf{LoadConfig} \\ 
    \textbf{Argument} & \textbf{Description} \\ 
    \textbf{batch\_size} & The number of rows to process in each batch (default: 131,072). \\ 
    \textbf{batch\_ readahead} & The number of batches to read ahead in a file (default: 16). By reading ahead, ParquetDB can reduce the waiting time when processing batches sequentially, as the next batch will already be in memory. However, increasing this value will require more RAM, so users need to balance performance and memory usage carefully, especially when working with large datasets. For smaller datasets, this may not be as critical, but for very large datasets, setting this value too high can cause memory exhaustion. \\ 
    \textbf{fragment\_ readahead} & The number of files to read ahead, improving IO utilization at the cost of RAM usage (default: 4). Similar to \code{batch\_readahead}, increasing this value can speed up I/O operations by keeping multiple files in memory at once, thereby reducing the delay when switching between files. However, more memory will be used as a result. This is particularly useful when dealing with large, partitioned datasets where data is spread across many files, such as in distributed systems or cloud storage environments. \\ 
    \textbf{fragment\_ scan\_options} & Options specific to a particular scan and fragment type, potentially changing across scans. \\ 
    \textbf{use\_threads} & Whether to use maximum parallelism determined by available CPU cores (default: \code{True}). \\ 
    \textbf{memory\_ pool} & The memory pool for allocations. Defaults to the system's default memory pool. \\ 
    \end{tblr}
    \end{adjustbox}

    \centering
    \begin{minipage}{\textwidth}
    \begin{lstlisting}
# Example usage:
from parquetdb import LoadConfig

config = LoadConfig(batch_size=65536, batch_readahead=8, fragment_readahead=2)
\end{lstlisting}
    \end{minipage}

    \label{table:load-config}
\end{table}

These parameters collectively give users more control over how data is loaded, allowing them to balance speed and memory usage depending on the size of the dataset and the available system resources. For large-scale datasets, tuning these options can lead to better performance and resource management.

\newpage

\subsubsection{Database Normalization}

Database normalization is essential for optimizing the performance of create, read, update, and delete (CRUD) operations in data management systems. By organizing data to minimize redundancy and maintain consistency, normalization ensures efficient data handling, particularly in large-scale datasets. Poorly structured data can result in uneven file distributions, excessive fragmentation, and inefficient access patterns, leading to performance bottlenecks. Normalization directly impacts CRUD operations by balancing row distribution across files, ensuring consistent row group sizes, and optimizing schema alignment. During data ingestion, normalization prevents excessive file creation and fragmentation, streamlining create operations. For reads, balanced file sizes and structured row groups enhance parallel processing, enable efficient predicate pushdown, and improve query performance. Updates benefit from minimized rewriting and predictable row alignment, while deletions are executed with reduced disruption to the dataset.

Normalization in ParquetDB can be done by calling the \code{normalize} function (Table \ref{table:normalize-function}). This function uses the \code{NormalizeConfig} class to configure the normalization process. More information about the \code{NormalizeConfig} class can be found in Table \ref{table:normalize-config}.

\begin{table}[h]
    \caption{Arguments and descriptions for the \code{normalize} function. Example code is listed below the table. Normalize the dataset by restructuring files for consistent row distribution. This method optimizes performance by ensuring that files in the dataset directory have a consistent number of rows. It first creates temporary files from the current dataset and rewrites them, ensuring that no file has significantly fewer rows than others, which can degrade performance. This is particularly useful after a large data ingestion, as it improves the efficiency of create, read, update, and delete operations.}
    \begin{adjustbox}{width=\textwidth,center}
    \centering
    \begin{tblr}{
        colspec = {X[0.15\textwidth,c,m]|X[0.75\textwidth,l,m]},
        column{1} = {valign=m, halign=c}, 
        column{2} = {halign=l}, 
        cell{2}{2} = {halign=c}, 
        cells = {font=\scriptsize\linespread{0.8}\selectfont},
        stretch = 0,
        rowsep = 3pt,
        hlines = {black, 1pt},
        vlines = {black, 1pt},
    }
    \SetCell[c=2]{c}\textbf{ParquetDB.normalize} \\ 
    \textbf{Argument} & \textbf{Description} \\ 
    \textbf{normalize\_ config} & Configuration for the normalization process, optimizing performance by managing row distribution and file structure. Accepts a \code{NormalizeConfig} object (optional). Default: \code{NormalizeConfig()}. \\
    \end{tblr}
    \end{adjustbox}

    \centering
    \begin{minipage}{\textwidth}

    \begin{lstlisting}
# Example usage:
from parquetdb import NormalizeConfig

normalize_config = NormalizeConfig(load_format='batches', 
                                   max_rows_per_file=5000, 
                                   max_rows_per_group=5000)

db.normalize(normalize_config=normalize_config)\end{lstlisting}
    \end{minipage}

    \label{table:normalize-function}
\end{table}


\begin{table}[htb]
    \caption{Arguments and descriptions for the \code{NormalizeConfig} class. This class provides configuration options for optimizing the normalization process, enabling efficient management of row distribution and file structures.}
    \begin{adjustbox}{width=\textwidth,center}
    \centering
    \begin{tblr}{
        colspec = {X[0.15\textwidth,c,m]|X[0.75\textwidth,l,m]},
        column{1} = {valign=m, halign=c}, 
        column{2} = {valign=m, halign=l},
        cell{2}{2} = {halign=c}, 
        cells = {font=\scriptsize\linespread{0.8}\selectfont},
        stretch = 0,
        rowsep = 3pt,
        hlines = {black, 1pt},
        vlines = {black, 1pt},
    }
    \SetCell[c=2]{c}\textbf{NormalizeConfig} \\ 
    \textbf{Argument} & \textbf{Description} \\ 
    \textbf{load\_format} & The format of the output dataset. Supported formats are \code{table} and \code{batches} (default: \code{table}). \\ 
    \textbf{batch\_size} & The number of rows to process in each batch (default: \code{None}). \\ 
    \textbf{batch\_ readahead} & The number of batches to read ahead in a file (default: 16). \\ 
    \textbf{fragment\_ readahead} & The number of files to read ahead, improving IO utilization at the cost of RAM usage (default: 4). \\ 
    \textbf{use\_threads} & Whether to use maximum parallelism determined by available CPU cores (default: \code{True}). \\ 
    \textbf{max\_ partitions} & Maximum number of partitions for dataset writing (default: 1024). \\ 
    \textbf{max\_ open\_files} & Maximum open files for dataset writing (default: 1024). \\ 
    \textbf{max\_rows\_ per\_file} & Maximum rows per file (default: 10,000). \\ 
    \textbf{min\_rows\_ per\_group} & Minimum rows per row group within each file (default: 0). \\ 
    \textbf{max\_rows\_ per\_group} & Maximum rows per row group within each file (default: 10,000). \\ 
    \end{tblr}
    \end{adjustbox}

    \centering
    \begin{minipage}{\textwidth}
    \begin{lstlisting}
# Example usage:
from parquetdb import NormalizeConfig

config = NormalizeConfig(load_format='batches', batch_size=500, max_rows_per_file=20000)
\end{lstlisting}
    \end{minipage}

    \label{table:normalize-config}
\end{table}

\newpage

\subsection{Using ParquetDB}


\begin{center}

\begin{minipage}{\textwidth}
\begin{lstlisting}
from parquetdb import ParquetDB

# Initialize the database
db = ParquetDB(db_path='parquetdb')


# Create data
data = [
    {'name': 'Alice', 'age': 30, 'occupation': 'Engineer'},
    {'name': 'Bob', 'age': 25, 'occupation': 'Data Scientist'}
]

# Add data to the database
db.create(data)

# Read data from the database
employees = db.read()
print(employees.to_pandas())


# Add another record with new data
data = [
    {'name': 'Jimmy', 'age': 30, 'state': 'West Virginia'},
]

# Add data to the database. The schema and current data will be updated to accomoodate for the state column
db.create(data)
\end{lstlisting}
\end{minipage}
\end{center}

\begin{center}
\begin{minipage}{\textwidth}
\begin{lstlisting}
# Read data from the database
employees = db.read()
print(employees.to_pandas())

# Currently Alice has null for state as she did not have it when her data was first input. Lets update this
update_data = [
    {'id':0, 'state': 'Maryland', 'zip':26709},
]

# Notice how all that is necessary is to get the id correponding to Alice. 
# Also we can provide addtional fields that do not exist in the dataset. 
# Since the field is new it will put null for the existing dataset
data.update(update_data)

# Read data from the database
employees = db.read()
print(employees.to_pandas())

# Deleting Jimmy. Since he was the thid to be put in his id is 2.
db.delete(id=[2])

# Read data from the database
employees = db.read()
print(employees.to_pandas())
\end{lstlisting}
\end{minipage}
\end{center}

\subsection{Comparison with other database}

\begin{table}[hbt]
    \begin{adjustbox}{width=\textwidth,center}
    \begin{tblr}{
        colspec = {X[0.20\textwidth,c,m]X[0.25\textwidth,c,m]X[0.25\textwidth,c,m]X[0.25\textwidth,c,m]},
        column{1} = {valign=m, halign=c},
        cells = {font=\footnotesize\linespread{0.8}\selectfont},
        stretch = 0,
        rowsep = 3pt,
        hlines = {black, 1pt},
        vlines = {black, 1pt},
      }
    \SetCell[c=4]{c}\textbf{Comparison of Database Systems} \\ 
    \textbf{Features} & \textbf{SQLite} & \textbf{MongoDB} & \textbf{ParquetDB} \\ 
    \textbf{ACID Compliance} & Full ACID compliance & Eventual consistency & Quasi-Durability (ACID) \\ 
    \textbf{Concurrency} & Limited (single writer; multiple readers) & High concurrency & Limited (single writer; multiple readers) \\ 
    \textbf{Portability} & Single file & Server-based & Single directory/file \\ 
    \textbf{Scalability} & Limited scalability (single machine) & Highly scalable (sharding; replication) & Limited scalability (single machine) \\ 
    \textbf{Complex Data Structures} & No & Yes & Yes \\ 
    \textbf{Schema Evolvability} & Fixed Schema & Dynamic Schema & Dynamic Schema \\ 
    \textbf{Encoding Techniques} & None & BSON encoding & Dictionary encoding; bit packing\\ 
    \textbf{Batching Support} & No native support & Supports batching with aggregation & Native Support \\ 
    \end{tblr}
    \end{adjustbox}
    \caption{Comparison of Database Systems: This table compares SQLite, MongoDB, and ParquetDB across various features.}
    \label{table:database-comparison}
\end{table}

To evaluate the strengths and limitations of ParquetDB, a comparative analysis with SQLite and MongoDB is presented in Table \ref{table:database-comparison}. This table provides a detailed assessment of features across the databases, including ACID compliance, concurrency, portability, support for complex data structures, schema evolvability, encoding techniques, batch processing capabilities, and compression methods. The most salient feature to emphasize is ACID compliance. ParquetDB upholds the principles of Atomicity, Consistency, and Isolation by generating temporary files prior to any modifications. For instance, when multiple updates are applied to a dataset, ParquetDB initially creates a duplicate of the data, implements the updates on this duplicate, and commits the changes only after all updates are successfully completed, thereby ensuring data consistency and isolation from incomplete modifications. In the event of an error during the modification process, these temporary files are reverted to maintain data integrity.

However, ParquetDB demonstrates partial adherence to the Durability principle. Specifically, it does not inherently recover from unexpected power failures or system crashes. While temporary files are utilized to safeguard ongoing changes, the recovery process necessitates manual intervention by a system administrator to restore these files and finalize the modifications. Consequently, during an incident such as a power outage or system crash, manual restoration of the temporary files by the user is essential to resume and complete the modification process.

\section{Benchmarking}

In this section, we outline the methodology used to evaluate the performance of ParquetDB in comparison to SQLite and MongoDB. The benchmarks focus on key performance metrics, including read, write, and update operations, and how these databases handle varying dataset sizes. Each subsection details specific aspects of the benchmarks, from setup and hardware specifications to the impact of data formats and indexing strategies. We also explore specialized scenarios, such as querying efficiency in large datasets and updating performance under real-world conditions. Through this structured approach, we aim to provide a comprehensive assessment of ParquetDB's strengths and limitations.

\subsection{Benchmark Setup}

To assess the performance of ParquetDB, we performed a series of benchmarks comparing it against SQLite and MongoDB. These benchmarks focused on read, write, and update operations across datasets of different sizes. The selection of hardware is crucial to ensure that the results are both reproducible and accurately reflect the system's computational capabilities and memory bandwidth. Understanding the hardware environment helps provide context for interpreting the benchmark results. The benchmarks were conducted on a system with the following specifications:

\begin{itemize}[label=-]
    \item \textbf{OS}: Windows 10 Pro
    \item \textbf{Processor}: AMD Ryzen 7 3700X 8-Core @ 3.6 MHz (8 cores, 16 logical processors)  
    \item \textbf{RAM}: 128 GB DDR4-3600 MHz (4x32 GB DIMMs) 
    \item \textbf{Storage}: SATA HDD 2TB (Model: ST2000DM008-2FR102)
\end{itemize}

Synthetic datasets consisting of 100 integer columns with varying record counts were used to simulate different load levels. The integer values ranged from 0 to one million, with dataset sizes varying from one row to one million rows. Random integers were chosen as they are a fundamental data type, allowing us to establish baseline performance metrics with minimal additional computational complexity. Integers were chosen as the benchmark data type since they are primitive and other data types can be extrapolated from their performance. Since integers are more lightweight compared to other data types, such as strings or nested structures, they provide an initial estimation of database performance without the variability introduced by more complex data representations. This approach allows us to evaluate the core efficiency of the database under ideal conditions, providing a best-case scenario. Subsequently, more complex data types may be expected to degrade performance due to increased memory usage and processing demands, and these results provide a reference point for such comparisons.

\subsection{Read and Write Performance}

The objective of this experiment was to evaluate read and write performance for varying numbers of records across SQLite, MongoDB, and ParquetDB. For write operations, the time required to insert records into each database and complete the transactions was measured. In the case of SQLite, the timing was taken prior to the executemany operation, whereas for MongoDB, it was measured both before using writemany and after the commit phase. Bulk insertions were employed for both databases, with optimizations for SQLite achieved using PRAGMA synchronous = OFF and PRAGMA journal mode = MEMORY. For read operations, the goal was to measure the time taken to load the entire dataset into an array-like structure, ensuring complete data retrieval without leaving any data in cursors.

\begin{figure}[ht]
    \centering
    \includegraphics[width=\textwidth]{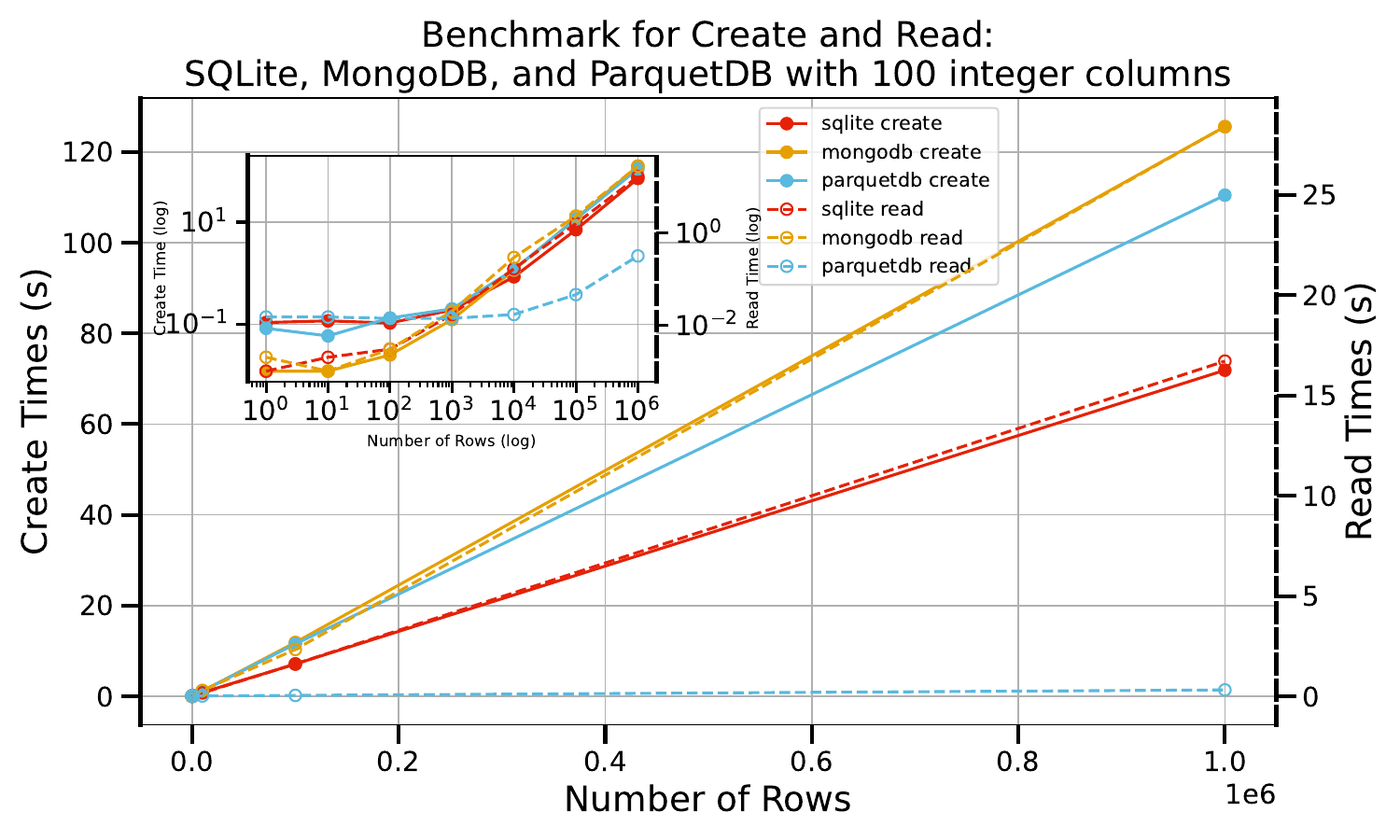}
    \caption{Benchmark Create and Read Times for Different Databases. Create time is plotted on the left y-axis, read time on the right y-axis, and the number of rows on the x-axis. A log plot is shown in the inset.}
    \label{fig:create-read-times}
\end{figure}

Figure \ref{fig:create-read-times} illustrates the create and read times for ParquetDB, SQLite, and MongoDB when working with a dataset comprising 100 integer columns. For smaller datasets, ParquetDB exhibits create times that are comparable to those of SQLite and MongoDB. However, as dataset size grows, ParquetDB demonstrates the second-best performance, following SQLite. In Figure \ref{fig:data-formats}, this performance can be be attributed to the data structure used in the tests; specifically, a Python list (pylist) was utilized, which suffers from suboptimal performance for creating large datasets due to its non-contiguous memory allocation. In terms of read performance, ParquetDB initially lags behind both SQLite and MongoDB for small datasets but shows considerable improvement as the dataset size increases, ultimately outperforming both competitors beyond a threshold of several hundred to a thousand rows. This improved performance can be largely attributed to the efficiency of Parquet's row-columnar storage format, which becomes increasingly advantageous as dataset size grows.

\subsection{Impact of Data Formats in ParquetDB}

We conducted a detailed evaluation of the impact of various data input formats on update performance in ParquetDB, specifically focusing on Python lists, Python dictionaries, Pandas DataFrames, and PyArrow Tables. These formats were chosen due to their common usage in Python-based data workflows, with Python lists and dictionaries being familiar to most users and Pandas DataFrames and PyArrow Tables offering compatibility with the backend library (PyArrow) used in ParquetDB. 

The experiment aimed to evaluate the update performance of ParquetDB using different data input formats. The key focus was on comparing the update times for varying numbers of rows, from small datasets to larger scales. The data formats tested include:

\begin{itemize}[label=-]
    \item Python lists (\code{pylist})
    \item Python dictionaries (\code{pydict})
    \item Pandas DataFrames (\code{pandas})
    \item PyArrow Tables (\code{pyarrow})
\end{itemize}

\begin{figure}[ht]
    \centering
    \includegraphics[width=\textwidth]{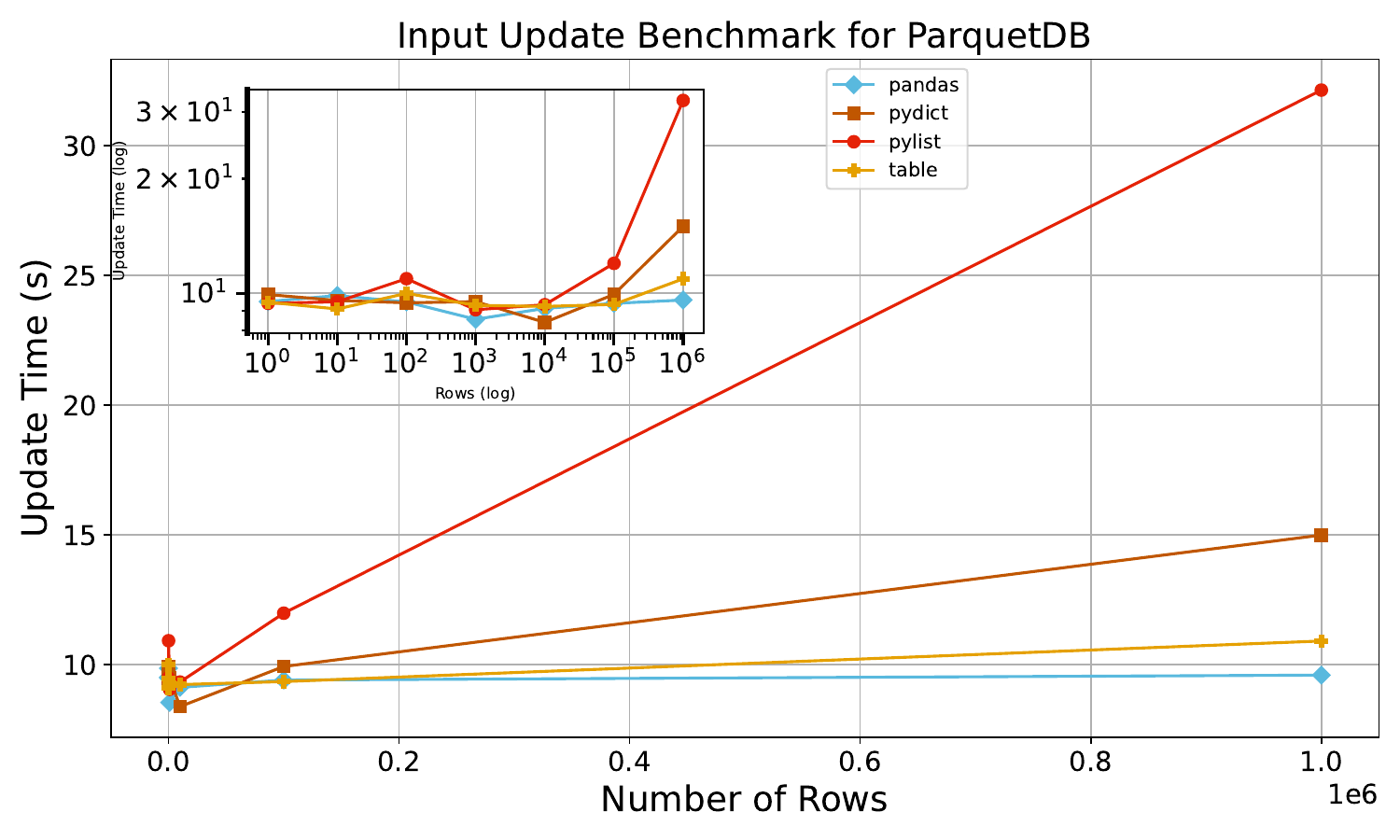}
    \caption{Update Time vs. Number of Rows for Different Data Formats in ParquetDB. Formats include Python lists, Python dictionaries, Pandas DataFrames, and PyArrow Tables. Log plot is shown in the inset.}
    \label{fig:data-formats}
\end{figure}

A plot of the update performance for each input format is presented in Figure \ref{fig:data-formats}. The plot provides a comparison of the time required to perform update operations as the number of rows increases, including an inset plot to better visualize the data on a logarithmic scale. For datasets with up to several thousand rows, the performance is relatively comparable across all input formats. However, as the size of the data set increases, significant divergences become apparent. 

PyArrow Tables and Pandas DataFrames exhibit the best performance, maintaining consistent update times even as the size of the dataset scales. This superior performance is largely due to the fact that ParquetDB relies on PyArrow to manage its internal data representation. This dependency results in native compatibility for PyArrow Table input. Moreover, Pandas DataFrames are well-supported by PyArrow, resulting in minimal data conversion overhead when interacting with ParquetDB.

Python dictionaries (pydict) show moderately inferior performance compared to PyArrow Tables and Pandas DataFrames. Although Python dictionaries store data types that are consistent (homogeneous) contiguously in memory, which can lead to improved data access patterns compared to non-homogeneous structures. However, they still require conversion to be fully compatible with PyArrow's internal representation, introducing additional processing overhead during updates.

Python lists (pylist) demonstrate the poorest performance as the dataset size grows. This is attributable to the non-contiguous memory allocation of Python lists, which necessitates extensive type conversions and leads to reduced memory efficiency when interfacing with PyArrow.

These benchmark results underscore that, for optimal update performance in ParquetDB, the use of PyArrow Tables or Pandas DataFrames is recommended, particularly for larger datasets. While Python lists and dictionaries are viable input types, they introduce non-trivial type conversion overhead, which adversely affects scalability and performance efficiency.In our benchmark experiments, we adopted a conservative approach by utilizing Python lists as the input format. This decision was intentional and aimed to establish a baseline performance level under suboptimal conditions, thereby accentuating the relative performance improvements achievable with more suitable data formats.

\subsection{Needle-in-a-Haystack Benchmark}

This experiment evaluates the performance of three databases: ParquetDB, SQLite, and MongoDB—in querying specific information from a field. The primary objective was to determine the efficiency with which each system could retrieve an arbitrary unique value inserted into a given column, reflecting scenarios where specific data points must be located quickly. Such operations are common in numerous real- world applications, such as searching for unique identifiers or specific records within large datasets. The dataset sizes varied, and the query times were measured for ParquetDB, SQLite, and MongoDB, with additional comparisons between indexed and non-indexed fields in SQLite and MongoDB.

\begin{figure}[ht]
    \centering
    \includegraphics[width=\textwidth]{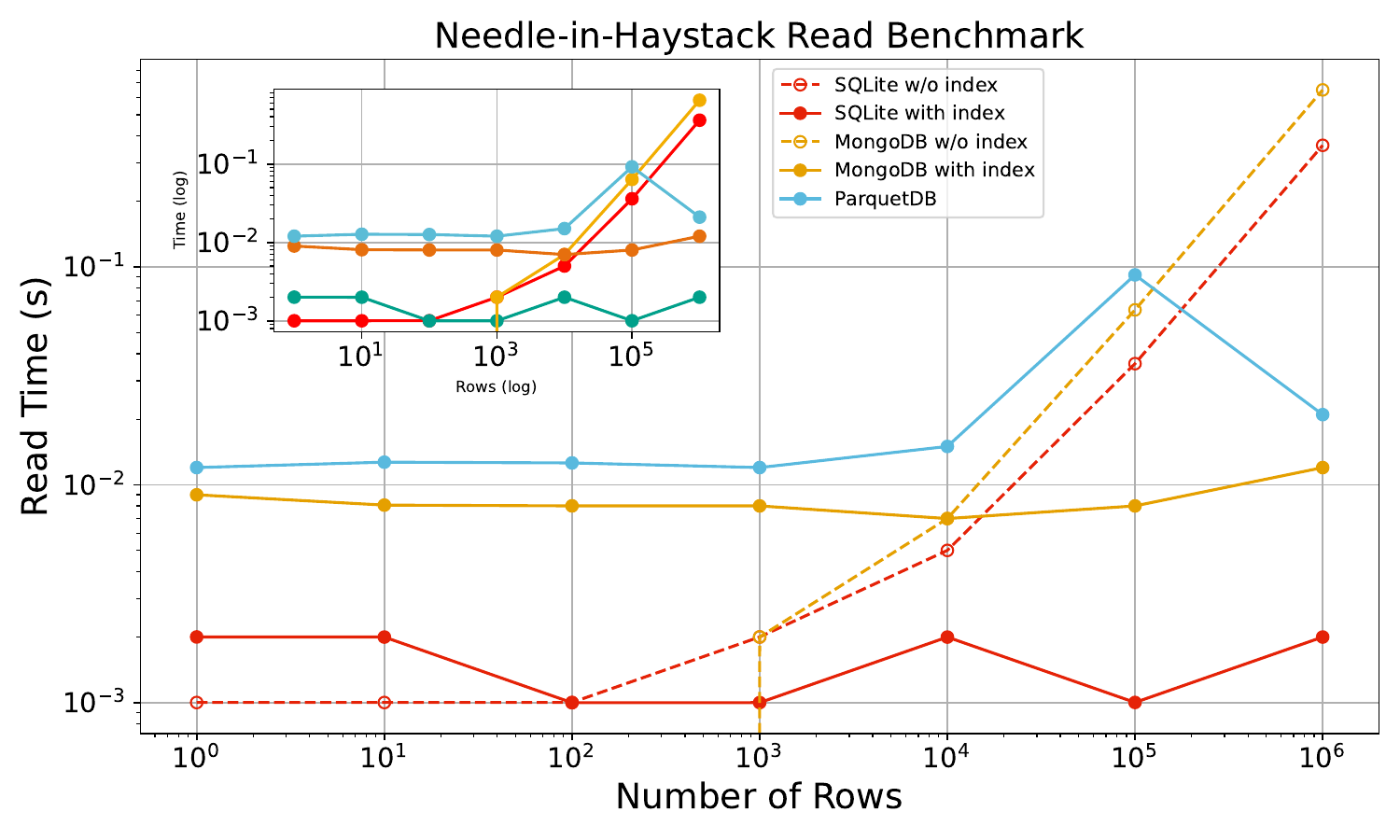}
    \caption{Needle-in-a-Haystack Benchmark Results. Time is on the y-axis, number of rows on the x-axis. The log plot is shown in the inset. SQLite and MongoDB are compared with and without indexing.}
    \label{fig:needle-in-haystack}
\end{figure}

Figure \ref{fig:needle-in-haystack} presents the results of this benchmark, highlighting the time required to retrieve specific records from each database. The plot follows a similar structure to previous performance evaluations, with query time represented on the y-axis and the number of rows on the x-axis. An inset logarithmic plot is included to improve the visualization of performance differences across varying scales.

For smaller datasets, ParquetDB exhibits significantly worse performance, lagging behind indexed SQLite, non-indexed SQLite, indexed MongoDB, and non-indexed MongoDB by approximately an order of magnitude. The relatively high query times in ParquetDB can be attributed to its lack of explicit indexing, resulting in a more exhaustive data scan compared to databases that leverage efficient indexing structures.

As dataset size increases, ParquetDB's performance improves significantly, ultimately becoming the third most efficient system, with a relatively constant query time as the number of rows grows. This trend of constant query time is also observed for indexed SQLite and indexed MongoDB, which can be attributed to the use of B-tree indexing. B-tree indexing is an efficient data structure that allows databases to maintain sorted data and perform searches, insertions, and deletions in logarithmic time, drastically reducing query time for indexed columns as dataset size scales. In contrast, the non-indexed SQLite and non-indexed MongoDB systems exhibit increasingly poor performance as dataset size grows, primarily due to their reliance on full table scans to locate the desired value. The absence of indexes forces these databases to perform linear scans, which results in longer query times with larger datasets.

Interestingly, ParquetDB is able to achieve query performance comparable to the indexed versions of SQLite and MongoDB, despite not utilizing traditional indexing mechanisms. This efficiency can be largely attributed to predicate pushdown filtering. Parquet files store field-level statistics in their schema, which allows for efficient filtering and retrieval of data without requiring a complete scan of the entire dataset. By leveraging these statistics, ParquetDB can effectively narrow down the search space, resulting in query times that are comparable to those of indexed systems.

The results of this experiment underscore the critical role of indexing in achieving efficient query performance, especially as the size of the dataset increases. Indexed SQLite and indexed MongoDB consistently demonstrate superior performance due to their B-tree indexing, which significantly reduces the complexity of search operations. Conversely, non-indexed versions of these databases suffer from substantial performance degradation as data scales. Despite the absence of traditional indexing, ParquetDB demonstrates competitive query performance, particularly for larger datasets. Its ability to leverage predicate pushdown and field-level statistics within Parquet files allows it to bypass the need for explicit indexes in many cases, thereby maintaining efficient query performance.

For optimal query performance, particularly with large datasets, indexed databases like SQLite and MongoDB are generally preferred. However, ParquetDB presents a compelling alternative, particularly for scenarios where indexing may not be feasible or desirable. By utilizing schema-level metadata and efficient filtering capabilities inherent in Parquet, ParquetDB can achieve performance close to that of indexed systems without incurring the additional overhead of index maintenance.

\subsection{Update Performance}

In this section, we evaluate the update performance of three databases, ParquetDB, SQLite, and MongoDB. The experiment involved preloading each database with one million rows to simulate a large-scale data environment commonly encountered in real-world applications. The objective was to measure the time required to update a varying number of records, thereby evaluating how each database handles bulk updates under conditions that approximate real-world workloads. To enhance update performance, an index was created on an ID column for both MongoDB and SQLite. Indexing allows databases to efficiently locate specific records that need updating, reducing the reliance on full table scans and significantly enhancing the overall efficiency of update operations.

\begin{figure}[ht]
    \centering
    \includegraphics[width=\textwidth]{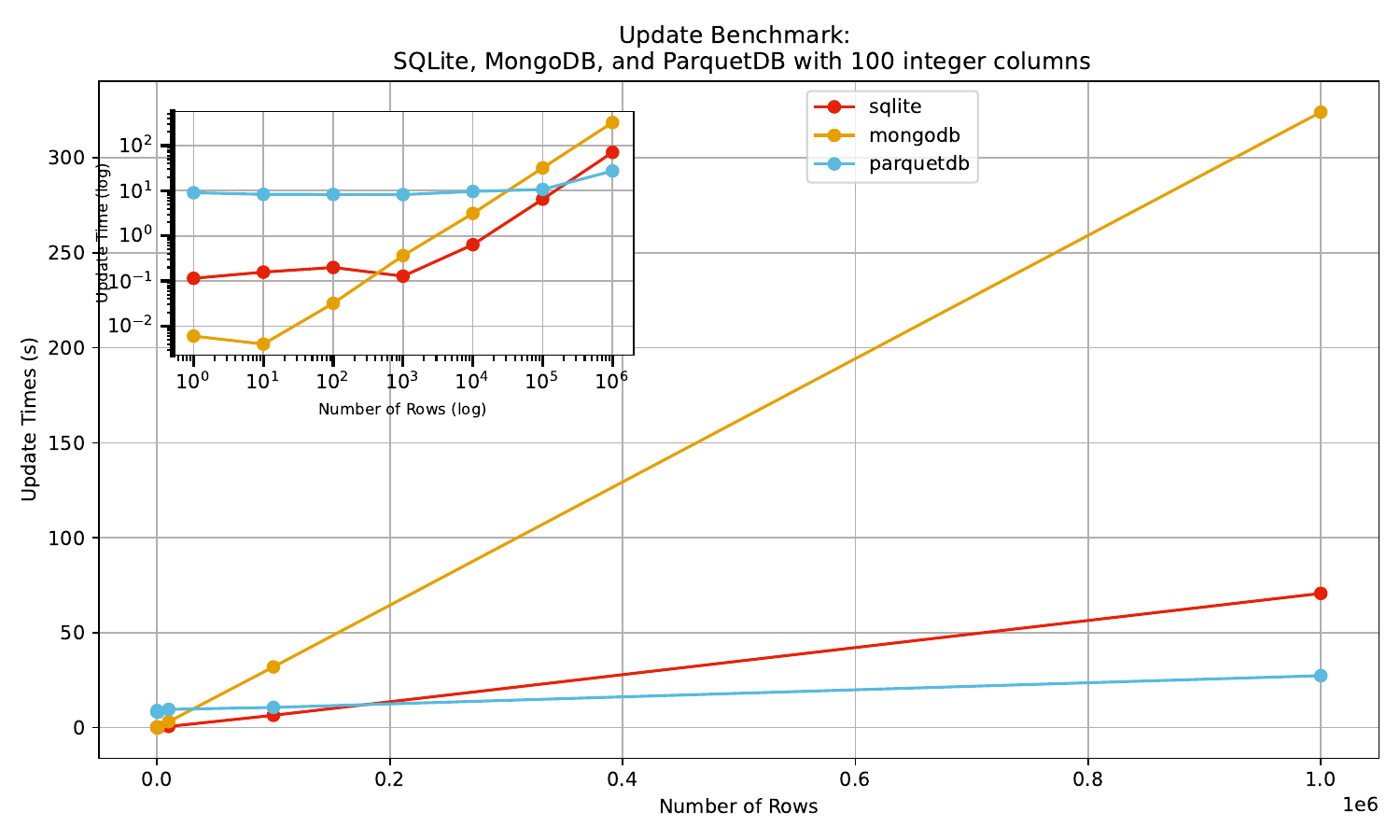}
    \caption{Update Times vs. Number of Rows for Different Databases. Update time is on the y-axis, number of rows on the x-axis, with a log plot in the inset.}
    \label{fig:update-performance}
\end{figure}

As shown in Figure \ref{fig:update-performance}, the results of the benchmark demonstrate how update times vary with the number of rows updated across the different databases. For smaller datasets, up to approximately 1,000 rows, ParquetDB exhibits the poorest performance compared to SQLite and MongoDB, with update times lagging behind by up to two orders of magnitude for tens of updated records. This performance deficiency can be attributed largely to the overhead associated with type conversion from Python lists (pylist) to formats compatible with PyArrow's columnar structure. As discussed in the earlier section, Python lists are not memory-contiguous, which necessitates significant conversion overhead, particularly affecting small datasets and resulting in delays.

\sloppy
Despite this initial disadvantage, ParquetDB's update performance demonstrates notable stability as the number of updated rows increases to approximately 1,000, underscoring its capacity to manage growing datasets efficiently. Beyond this point, update times start to increase, yet ParquetDB continues to exhibit competitive performance relative to the other databases. When updating a substantial number of records, ParquetDB ranks as the best performer. For instance, when updating one million rows, ParquetDB takes approximately 27 seconds. Notably, of the 27 seconds taken by ParquetDB, around 18 seconds are attributed to type conversion. Although MongoDB is competitive for smaller update operations, it falls behind SQLite and ParquetDB as the number of updated rows increases. This observation highlights the efficiency of ParquetDB in scaling with larger datasets, despite the type conversion bottleneck.

The experiment's findings reveal that ParquetDB can compete effectively with traditional databases like SQLite and MongoDB for update operations, particularly as dataset size increases. While SQLite demonstrates superior update times due to its optimized indexing and internal mechanisms, ParquetDB performs well, especially considering its use of a general-purpose data format and the additional overhead of type conversion. ParquetDB's scalability is particularly noteworthy, maintaining efficient update times even with growing dataset sizes. ParquetDB remains competitive, scaling effectively with larger datasets despite inherent type conversion costs. MongoDB, initially competitive, struggles to maintain performance for larger updates compared to the other databases.

\section{Real-World Application: Alexandria 3D Materials Database Application}

In this section, we evaluate the performance of ParquetDB in managing a real-world dataset, focusing on key performance metrics such as data load time, input time, and execution efficiency for database operations. Specifically, we use the Alexandria 3D Materials Database, a comprehensive dataset comprising approximately 4.3 million unique material structures \cite{schmidtMachineLearningAssistedDeterminationGlobal2023}. Our evaluation involves measuring the time required to load data from JSON files, input these records into ParquetDB, and execute various operations on the database. These metrics provide critical insights into the efficiency and viability of employing ParquetDB for large-scale data management tasks.

The Alexandria 3D Materials Database was developed to address inherent biases present in previous materials databases, particularly the overrepresentation of common elements such as silicon and oxygen, as well as limited diversity in crystal structures. These biases often resulted in skewed predictions and reduced generalizability of machine learning models trained on these datasets. To overcome these limitations, the Alexandria dataset was generated using a large-scale, stepwise approach that integrated high-throughput density functional theory (DFT) calculations with advanced machine learning techniques, including transfer learning.

Initially, a broad chemical space of 84 elements was explored, including lanthanides and some actinides, to ensure comprehensive chemical diversity. The database includes a wide range of crystal structures, such as ternary garnets, Ruddlesden-Popper layered perovskites, cubic Laves phases, and Heusler compounds. A crystal graph attention network (CGAT) was employed to predict compound stability, followed by iterative DFT calculations to verify these predictions. This iterative process allowed the dataset to grow significantly, covering a much broader and diverse set of compounds compared to its predecessor, DCGAT-1\cite{schmidtLargescaleMachinelearningassistedExploration2022}.

Our package provides a method to automatically download the dataset:

\begin{lstlisting}
parquetdb.external_utils.download_alexandria_3d_database(output_dir)
\end{lstlisting}

\noindent and an accompanying script (\code{Example 1 - 3D Alexandria Database.py}) to execute this experiment. The dataset is provided in compressed JSON files, which are downloaded and decompressed into a specified directory. Each JSON file contains $100,000$ records, except for the final file, which contains $89,295$ records. The data is in the form of a list of dictionaries. Each record is represented as a nested dictionary, with keys including \code{@class} (string), \code{@module} (string), \code{composition} (dictionary), \code{data} (nested dictionary), \code{energy} (float), \code{energy\_adjustments} (list of dictionaries), \code{entry\_id} (string), \code{parameters} (dictionary), and \code{structure} (nested dictionary representing a pymatgen Structure object).

\subsection{Loading Data}

\begin{figure}[ht]
    \centering
    \includegraphics[width=\textwidth]{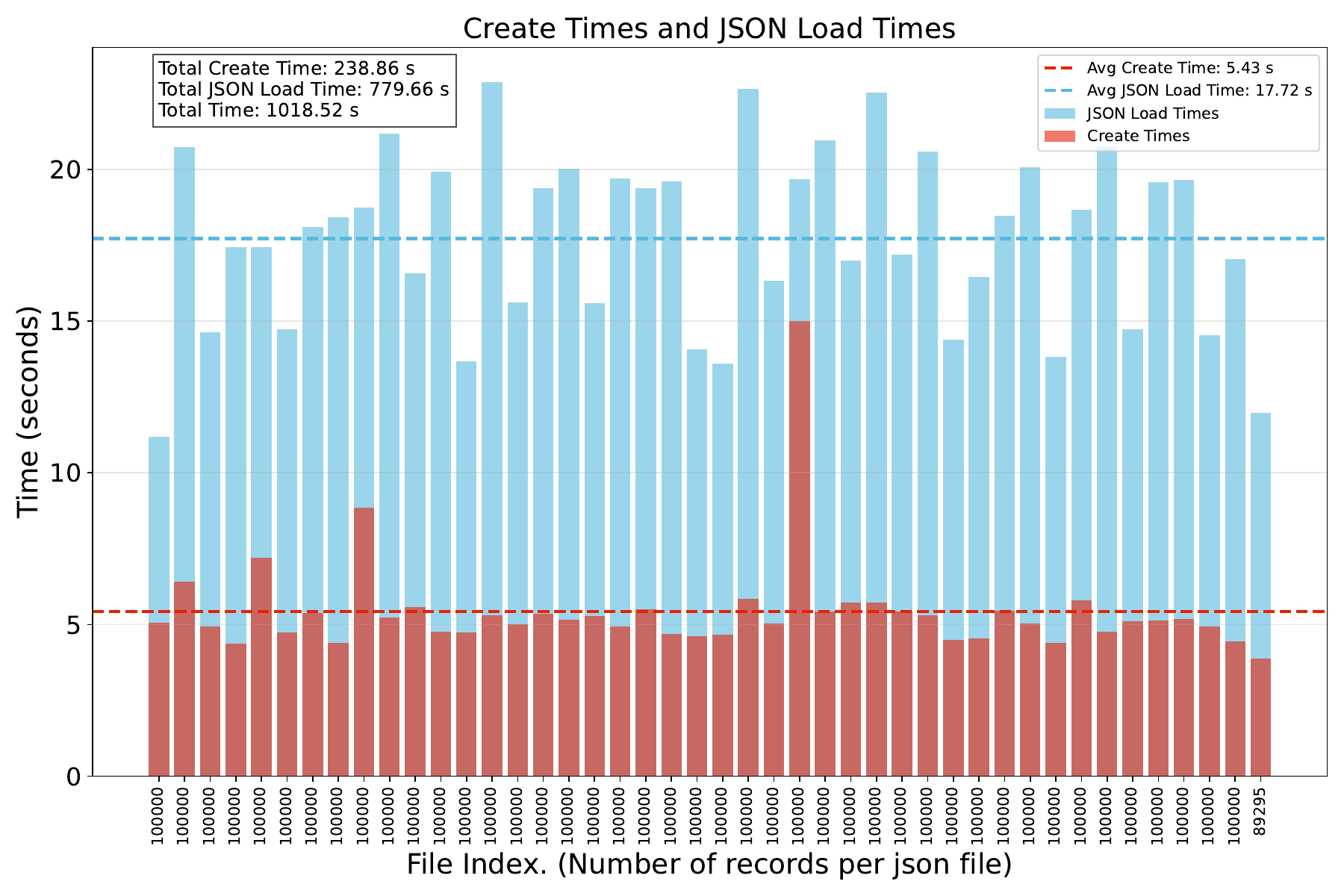}
    \caption{Create and JSON Loading Time for the Alexandria 3D Materials Database. This plot shows the time to load the json file (blue) and the time to create the records in ParquetDB (red).}
    \label{fig:alexandria-create}
\end{figure}

The results of the loading process are depicted in Figure \ref{fig:alexandria-create}, which presents a bar plot of the load and create times for all JSON files. Most JSON files contain $100,000$ records, except for the final one with $89,295$ records. The code to insert each batch is:

\begin{lstlisting}
    with open(json_file, 'r') as f:
        data = json.load(f)
    db.create(data=data['entries'])
\end{lstlisting}

\noindent The load times for the JSON files are shown in blue, while the create times are shown in red. Loading the JSON files took an average of $17.72$ seconds per file, with a cumulative time of $727$ seconds. This step accounts for the majority of the upload time, as expected due to the need to deserialize the text-based JSON format into Python objects. Creating records in ParquetDB took an average of $5.60$ seconds per file, resulting in a total time of $238$ seconds for the entire dataset. This performance is significant compared to similar operations in databases like SQLite or MongoDB, where inserting records of similar complexity can be considerably more time-consuming due to limitations in handling nested structures and schema evolution. For example, SQLite requires either flattening nested dictionaries or creating separate relational tables, both of which introduce additional computational overhead. Writing $4.3$ million records to the database in $246$ seconds. Some files exhibited longer creation times due to schema changes, which requires updating the entire database to maintain schema consistency.



\subsection{Additional Evaluations}

\begin{figure}[ht]
    \centering
    \includegraphics[width=\textwidth]{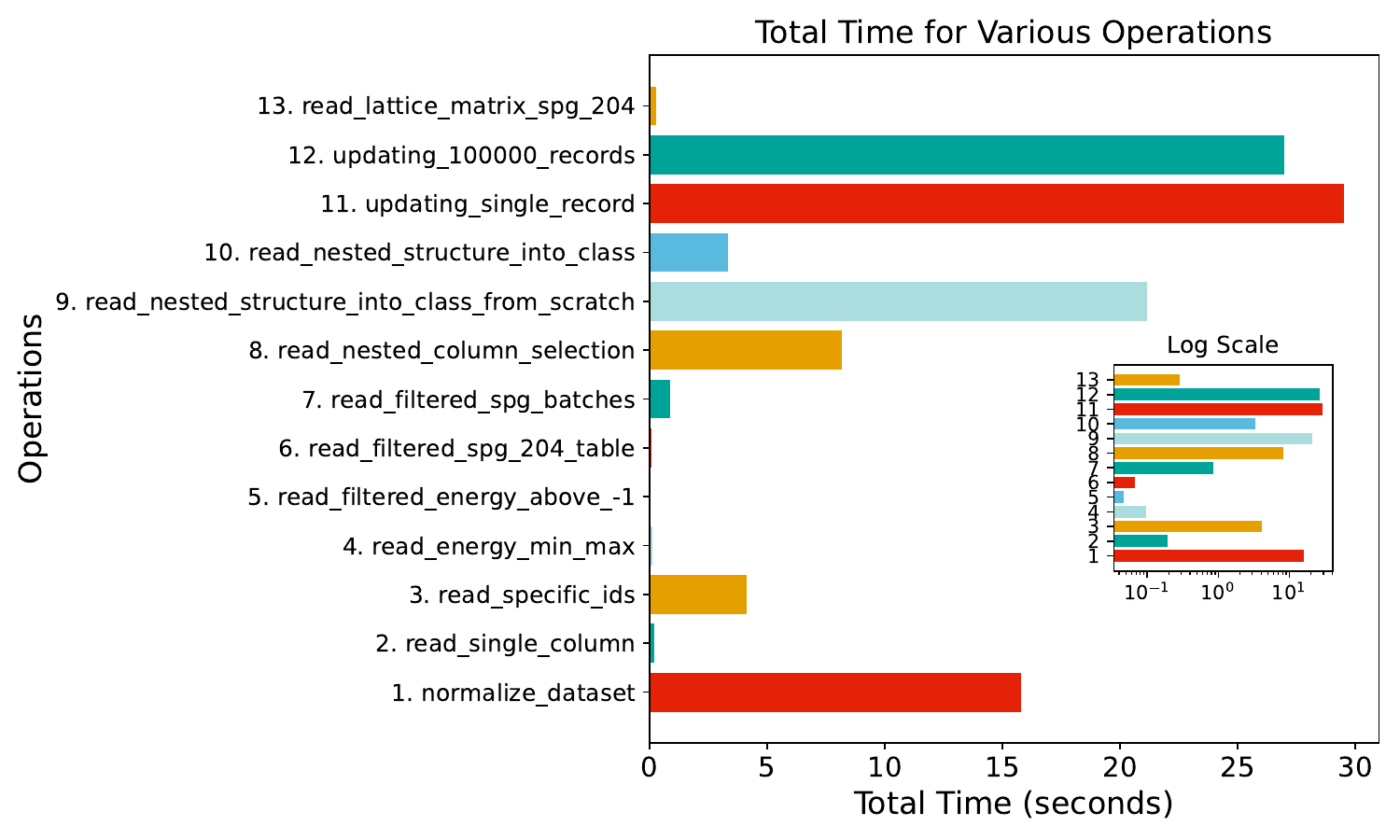}
    \caption{Performance of Various Operations on the Alexandria 3D Materials Database. A horizontal bar chart is presented, with a log plot in the inset.}
    \label{fig:alexandria-operations}
\end{figure}

We conducted additional evaluations to benchmark the querying and performance capabilities of ParquetDB The results are presented in Figure \ref{fig:alexandria-operations}, which includes a horizontal bar plot depicting the time required to complete various tasks, with an inset showing results on a logarithmic scale. he subsequent sections detail the implementation of each operation, supported by relevant code snippets to facilitate reproducibility and deeper understanding.

\subsubsection{Database Normalization} 

\begin{lstlisting}
db.normalize(
batch_size = 100000,          
max_rows_per_file = 500000,  
max_rows_per_group = 500000
)
\end{lstlisting}

The process of database normalization in ParquetDB aims to optimize storage and improve operational efficiency by consolidating smaller files and removing redundant data. This approach reduces the total number of files, thereby minimizing input/output (I/O) operations, which are often a bottleneck in high-performance data workflows. By grouping records into larger, well-structured files, ParquetDB enhances both read and write performance, especially for large datasets. The normalization process for the evaluated dataset completed in 17.4 seconds.

\subsubsection{Reading a Single Column} 

\begin{lstlisting}
table = db.read(columns=['id'])
\end{lstlisting}
In this example, we only read the \textbf{id} column, which contains all 4.3 million IDs, into memory.This operation completed in approximately 0.1 seconds, showcasing the speed and scalability of ParquetDB. This demonstrates the efficiency of column projection pushdown, wherein ParquetDB allows the selection of specific columns without the need to load the entire dataset into memory.

\subsubsection{Querying 10 Specific IDs} 

\begin{lstlisting}
table = db.read(ids=[0, 10, 100, 1000, 10000, 100000, 1000000])
\end{lstlisting}
In this example, we queried 10 specific IDs and read all columns into memory, which took approximately 3 seconds to complete.

\subsubsection{Reading Energy Values and Calculating Extremes} 

\begin{lstlisting}
table = db.read(columns=['energy']);
result = pc.min_max(table['energy']);
min_value = result['min'].as_py();
max_value = result['max'].as_py();
\end{lstlisting}
In this example, we test how long does it take to perform operations on the energy column. Again, we select only the energy column to read into memory, then PyArrow has compute functions to fully take advantage of parallelism. This operation  took approximately 0.1 seconds. 

\subsubsection{Filtering Energies Above -1 eV} 

\begin{lstlisting}
db.read(columns=['id', 'energy'], filters=[pc.field('energy') > -1.0])
\end{lstlisting}
In this example, we retrieved rows where the energy is greater than -1 eV, taking approximately 0.05 seconds. This highlights how predicate pushdown can be used to screen the rows in ParquetDB. Column projection pushdown may also be used as well to reduce memory usage. In other databases, achieving comparable performance often necessitates indexing, which adds overhead. ParquetDB, on the other hand, provides efficient querying without additional indexing through the parquet's metadata.

\subsubsection{Querying Structures by Space Group} 
\begin{sloppypar}
\begin{lstlisting}
table = db.read(columns=['id', 'data.spg'], filters=[pc.field('data.spg') == 204])
\end{lstlisting}
In this example, we tested how long it takes to filter the space group. This is a similar illustration to the energy example, however, the space group lies in a nested field. This took around 0.1 seconds. This example demonstrates how filter nested structures, using the syntax \code{\{parent\}.\{child1\}.\{child2\}} to the reference fields.
\end{sloppypar}

\subsubsection{Batch Querying by Space Group} 

\begin{lstlisting}
generator = db.read(
    columns=['id', 'data.spg'], 
    filters=[pc.field('data.spg') == 204], 
    load_format='batches', 
    batch_size=1000)
\end{lstlisting}
In this example, we show how to read the data in batches. This takes approximately 0.02 seconds. Instead of returning an in-memory table, we obtained a generator for iterating over the results. This approach is advantageous for processing large datasets, as it allows data to be handled in smaller, memory-efficient chunks, thereby reducing memory usage and enhancing overall performance. The batch-based approach also yielded significant performance improvements since filtering was applied on individual batches, allowing empty batches to be skipped during iteration.

\subsubsection{}{Reading Nested Subfields as Lists of Dictionaries} 

\begin{lstlisting}
table = db.read(columns=['id', 'structure.sites'])
\end{lstlisting}
In this example, we show how to access nested structures (\code{structure.sites}), which contains lists of dictionaries detailing the atomic positions within materials. To access the nested structure the following pattern must be used: \code{\{parent\}.\{child1\}.\{child2\}} to reference the fields. Since these lists can be quite large, with up to 100 atomic sites per material, the read time was approximately 2.5 seconds, which remains efficient given the magnitude of data—reading atomic site information for 4.8 million structures.

\subsubsection{Rebuilding Nested Structures into Class From Scratch} 

To read and reconstruct nested structures, we used:
\begin{lstlisting}
table = db.read(
    columns=['id', 'structure', 'data'], 
    ids=[0], 
    rebuild_nested_struct=True, 
    rebuild_nested_from_scratch=True)
\end{lstlisting}

By default, nested data is flattened during ingestion to optimize read and write performance. However, scenarios often arise where it is necessary to reconstruct these nested structures in their original hierarchical form. This example demonstrates this reconstruction process, rebuilding the nested structure for a single record. This operation took approximately 23 seconds to complete, as the system reconstructs the entire nested structure from scratch. The longer execution time reflects the computational complexity of reorganizing flattened data into its original format. However, this functionality is invaluable when the original hierarchical data is required, such as using a nested structure dictionary to initialize objects like pymatgen.core.Structure, which are commonly used in materials science workflows.

The upfront cost of reconstructing nested structures can be justified by the subsequent performance benefits. Once the data has been reconstructed, subsequent reads do not require rebuilding, leading to significantly faster execution times, as demonstrated in the next section.

\subsubsection{Rebuilding Nested Structures into Class} 

\begin{lstlisting}
table = db.read(
    columns=['id', 'structure', 'data'], 
    ids=[0], 
    rebuild_nested_struct=True, 
    rebuild_nested_from_scratch=False)
\end{lstlisting}

In this example, we test how long it takes to read when ParquetDB has already reconstructed the nested structures. This operation was completed in approximately 3 seconds, a significant improvement compared to the 23 seconds required for the initial reconstruction.

\subsubsection{Updating a Single Record} 

\begin{lstlisting}
db.update([{'id':0, 'data.spg':210}], 
    normalize_config=NormalizeConfig(
        load_format='batches',
        batch_readahead=10,
        fragment_readahead=2,
        batch_size = 100000,
        max_rows_per_file=500000,                                                 
        max_rows_per_group=500000)
  )
\end{lstlisting}

In this operation, a single record with \code{id=0} is updated by modifying its \code{data.spg} value to 210. The update process incorporates normalization settings, which ensure that data remains compact and optimized for subsequent operations. This update took approximately 30 seconds, reflecting the comprehensive reorganization required to maintain the file and group structures within the Parquet format.

While the time required may seem substantial for a single update, this overhead is largely due to the lack of native in-place updates in Parquet files, which are inherently designed for high-throughput read operations. Each update triggers a rewriting process to ensure data integrity and compatibility with subsequent queries.

\subsubsection{Updating 100,000 Records} 

\begin{lstlisting}
updates = [{'id': np.random.randint(0, table.num_rows), 'data.spg': 210} for _ in range(100000)]
db.update(updates, 
    normalize_config=NormalizeConfig(...)
  )
\end{lstlisting}

The following example shows how 100,000 records are updated. This batch update operation completed in approximately 25 seconds, outperforming the single-record update despite its significantly larger scope. The reduced time per record can be attributed to batch optimization techniques, which minimize repetitive file access and I/O operations by processing updates in groups.

\subsubsection{Read Nd Lattice Matrix} 

\begin{lstlisting}
updates = [{'id': np.random.randint(0, table.num_rows), 'data.spg': 210} for _ in range(100000)]
table = db.read(columns=['structure.lattice.matrix'], 
                    filters=[pc.field('data.spg') == 204])

lattice = table['structure.lattice.matrix'].combine_chunks().to_numpy_ndarray()
\end{lstlisting}

The following example demonstrates how to read a 2D lattice matrix from the dataset. In this operation, the \code{structure.lattice.matrix} field, representing a nested array of lattice parameters, is read into memory and converted into a NumPy array for further computation. The filtering step ensures that only records with a space group of 204 are included in the result, optimizing both memory usage and processing time. The entire operation completed in approximately 0.1 seconds.

\subsubsection{Determining Electrical Property Distrbution}

In this example, we demonstrate how to use ParquetDB to query the Alexandria3D database for material distribution based on electrical properties. First, we categorize materials into metals, small gap materials, semiconductors, and insulators, based on their energy band gap ($E_g$). The energy band gap defines the separation between the valence (occupied) and conduction (unoccupied) bands. Metals have a band gap of 0 eV, allowing electrons to move freely. Insulators have a large band gap (>3 eV), preventing electron movement. Semiconductors have a moderate band gap (0.1 - 3 eV), allowing electron movement with small external stimuli. Small gap materials (0 - 0.1 eV) allow thermal excitation to facilitate electron movement.

The Alexandria3D database records both direct and indirect band gaps. The direct band gap is the energy difference between the top of the valence and bottom of the conduction band (Gamma point). The indirect band gap is the difference between the maximum valence and minimum conduction band energies. We classify materials based on the smaller non-zero value.

\begin{figure}[pht]
    \centering

    \begin{subfigure}[b]{0.7\textwidth}
        \includegraphics[width=\textwidth]{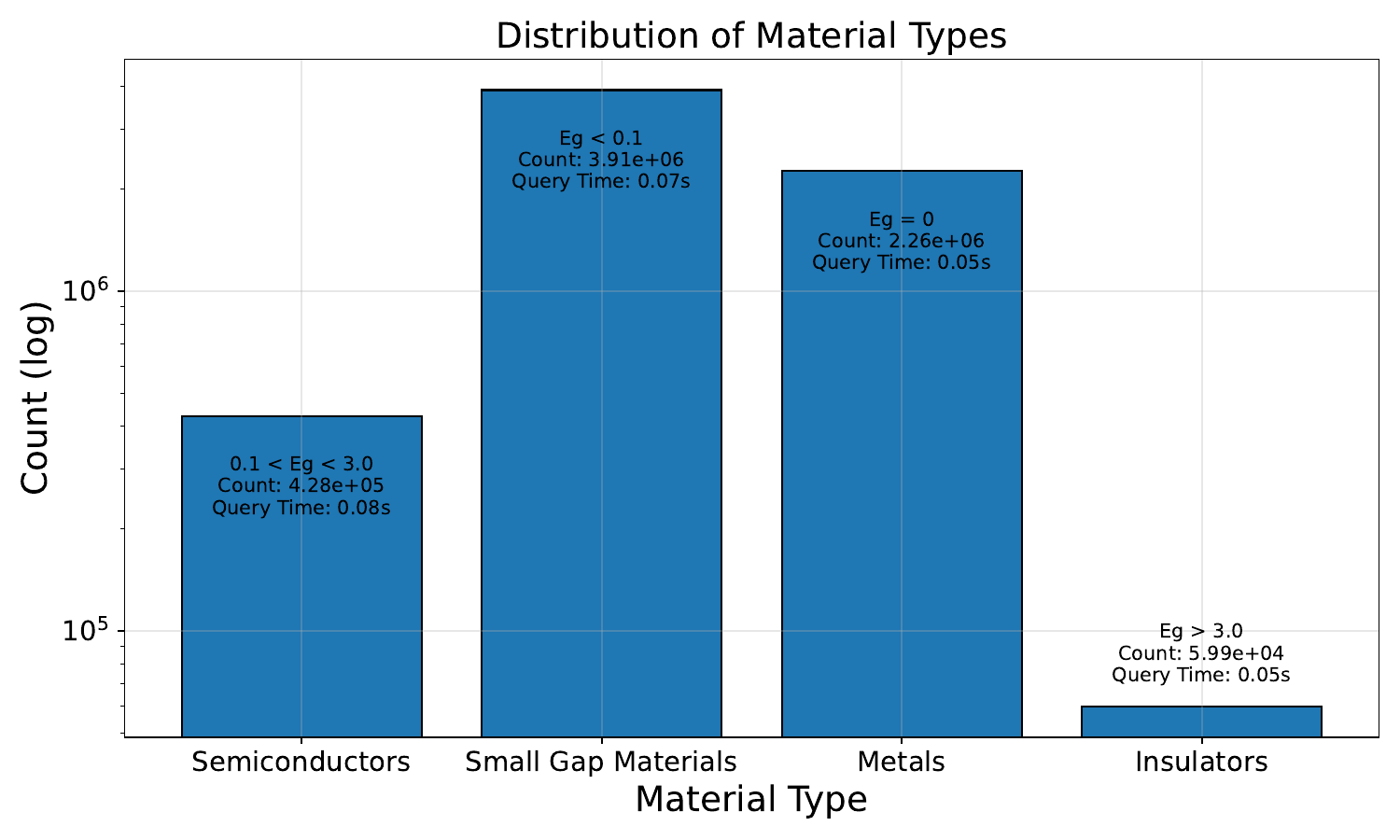}
        \caption{}
        \label{fig:alexandria-electrical-properties-dist}
    \end{subfigure}

    \begin{subfigure}[b]{0.8\textwidth}
        \includegraphics[width=\textwidth]{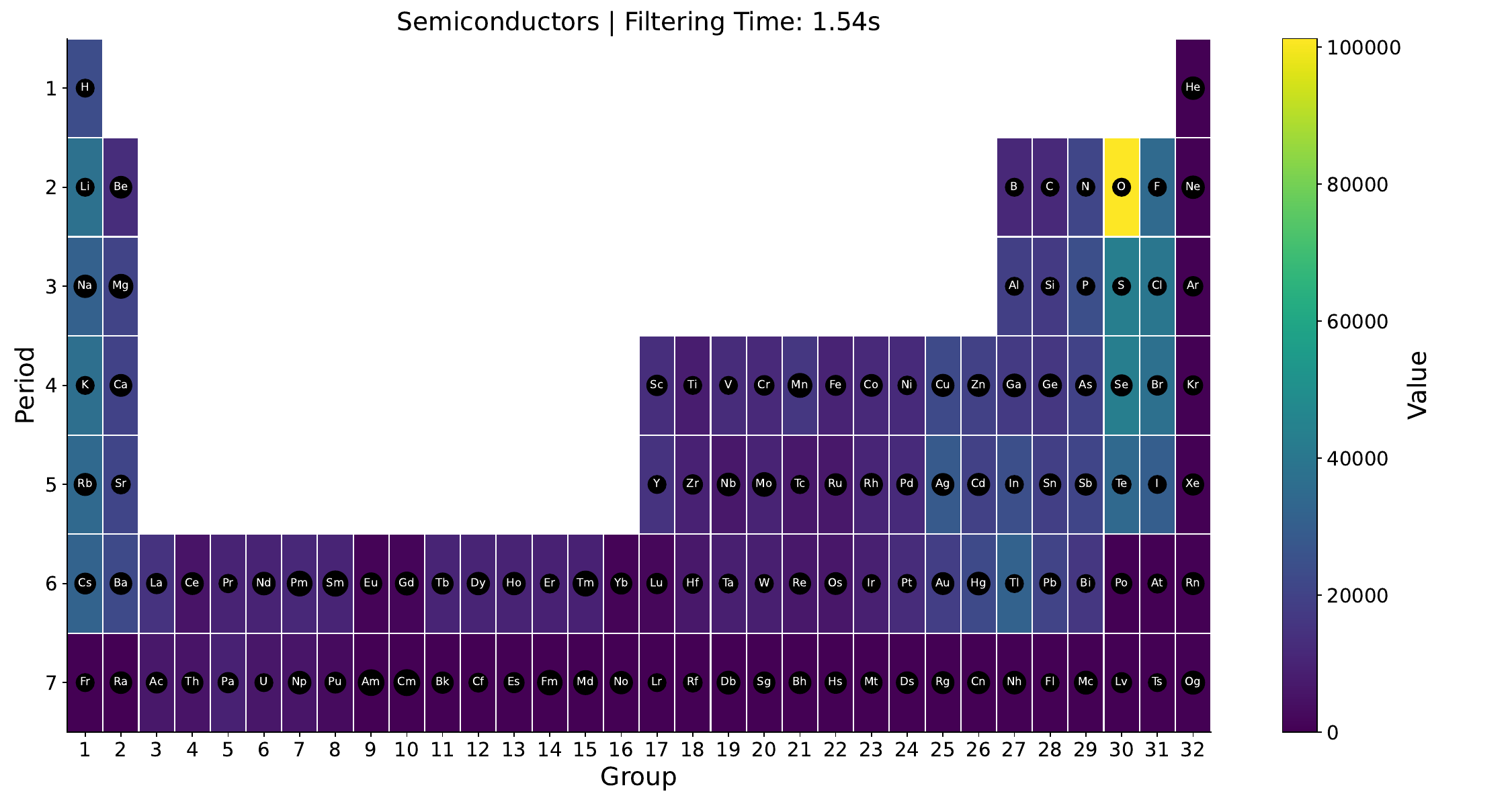}
        \caption{}
        \label{fig:alexandria-electrical-properties-periodic}
    \end{subfigure}
    \caption{Analysis of electrical properties in the Alexandria3D database. (\ref{fig:alexandria-electrical-properties-dist}) Distribution of materials in the Alexandria3D database based on their electrical properties. The materials are classified as metals ($E_g$ = 0 eV), small gap materials ($0 < E_g < 0.1$ eV), semiconductors ($0.1 < E_g < 3$ eV), and insulators ($E_g > 3$ eV). (\ref{fig:alexandria-electrical-properties-periodic}) Periodic table heatmap showing the distribution of semiconductor materials in the Alexandria3D database. The color intensity indicates the frequency of occurrence of each element in semiconductor compounds.}
    \label{fig:alexandria-electrical-properties}

\end{figure}

We use PyArrow's expression language to create complex queries. Here, we filter materials based on whether the indirect band gap is non-zero and smaller than the direct band gap. If true, we filter using the indirect band gap; otherwise, we use the direct band gap. The results are visualized in Figure \ref{fig:alexandria-electrical-properties-dist}, showing the distribution of materials. Insulators are the least common (~59,900 records), followed by semiconductors (~428,000 records), while metals (~2.26 million) and small gap materials (~3.91 million) are more prevalent.

\begin{center}
\begin{minipage}{\textwidth}
\begin{lstlisting}
# Semiconductor
filtered_expr = pc.if_else(
(pc.field("data.band_gap_ind") != 0) & (pc.field("data.band_gap_ind") < pc.field("data.band_gap_dir")),
(pc.field("data.band_gap_ind") > 0.1) & (pc.field("data.band_gap_ind") < 3),
(pc.field("data.band_gap_dir") > 0.1) & (pc.field("data.band_gap_dir") < 3)
)
semiconductor_table=db.read(columns=['id'], filters=[filtered_expr])
\end{lstlisting}
\end{minipage}
\end{center}

To further analyze semiconductors, we determine the periodic table distribution of their elements. The 'data.elements' property in Alexandria3D provides the list of elements for each material. Due to limited list filtering options in PyArrow, we use a manual approach to count element occurrences. Figure \ref{fig:alexandria-electrical-properties-periodic} shows the periodic table distribution of elements in semiconductors, aligning with expectations. Transition and rare-earth metals are less common, as they typically form metallic bonds, while elements from groups 1-2 and 27-31 are more common, forming ionic and covalent bonds that localize electrons.

\begin{lstlisting}
    for key in values_element_dict.keys():
        flat_list = pc.list_flatten(semiconductor_table['data.elements'])
        flat_list_indices = pc.list_parent_indices(semiconductor_table['data.elements'])
    
        equal_mask = pc.equal(flat_list, value)
        equal_table_indices = pc.filter(flat_list_indices, equal_mask)
        filtered_table = pc.take(table, equal_table_indices)
        number_of_occurrences = filtered_table.shape[0]
        values_element_dict[key] = number_of_occurrences
\end{lstlisting}

The query times across all material types are approximately equivalent, taking around 0.06 seconds. The filtering time required to generate the periodic table distribution is approximately 1.12 seconds. These results should be compared with other data storage formats discussed in the paper. Although we have not conducted the same analysis for these formats, we can estimate completion times and address implementation complexity based on benchmarks and previous examples. Understanding the nuances of each format allows us to appreciate the benefits and limitations inherent to different data storage solutions, providing a more comprehensive view of their applicability in various scenarios.

For JSON, the estimated query time would be at least 779.66 seconds, primarily due to the necessity of reading the entire dataset into memory, unless one employs an exceptionally sophisticated approach for the conditional query. JSON, being a text-based, human-readable format, is inherently less efficient for complex querying tasks, particularly when compared to other formats optimized for data storage and retrieval. This inefficiency is exacerbated when dealing with large datasets, where the absence of inherent indexing and the need to traverse the entire data structure significantly impede performance. Consequently, JSON is better suited for data interchange rather than heavy analytical querying, as highlighted by the substantial performance lag observed in our estimations.

In the case of SQLite and MongoDB, the query performance would depend on whether the direct and indirect band gaps are indexed. If indexed, we could expect query times to be competitive, potentially slightly faster than ParquetDB. However, there is a significant increase in implementation complexity. One must first create appropriate indexes on the direct and indirect band gap fields and then craft a SQL query capable of executing the desired conditional filtering. The process of index creation, while beneficial for optimizing query performance, introduces an overhead in terms of both time and resource management. Moreover, even with indexes in place, the retrieval process may require additional computational resources to maintain and update these indexes, especially as the dataset grows or evolves over time. Even after the initial retrieval of data, further filtering would require additional processing efforts, which adds layers of complexity to the workflow. 

In contrast, ParquetDB offers a more straightforward implementation. The data is directly returned in a PyArrow Table, which facilitates subsequent filtering and operations seamlessly, as demonstrated in the generation of the periodic table distribution. This streamlined approach reduces both the implementation complexity and the overhead associated with additional data manipulation. By utilizing a columnar storage format, ParquetDB allows for efficient read access to specific columns without needing to scan the entire dataset, thereby significantly optimizing performance for analytical queries. Additionally, PyArrow's native integration with Parquet files ensures that the filtering and transformation steps are performed in-memory with minimal latency, further improving the efficiency of data operations.

\section{Conclusion and Future Work}

In this paper, we introduced ParquetDB, a Python-based database framework that leverages the Parquet file format to address the limitations inherent in traditional data storage formats and database systems. ParquetDB presents a compelling solution for scenarios demanding efficient data retrieval, low complexity in implementation, and scalability.

The need for efficient, scalable, and accessible data storage solutions has never been more critical in an era where data underpins technological advancement and research. As we push the boundaries of artificial intelligence and data-driven decision-making, the tools we use to manage, store, and retrieve data must be able to adapt to these fast-paced changes. Traditional data formats such as JSON, CSV, and TXT, alongside database management systems like SQLite and MongoDB, have long served as foundational solutions for data handling in various domains. However, they often introduce complexities and inefficiencies that limit rapid iteration and experimentation—elements crucial to cutting-edge research and development.

One of the key distinctions of ParquetDB lies in its ability to overcome the serialization and deserialization inefficiencies found in traditional data formats. As discussed in this paper, the Parquet file format's columnar structure allows for more efficient encoding of numerical data, which significantly reduces both storage space and data retrieval times compared to text-based formats like JSON and CSV. This optimization is particularly crucial when handling large datasets, where performance can degrade significantly due to the inherent limitations in traditional formats that rely on plain text encoding. ParquetDB's design mitigates these issues by utilizing advanced compression techniques and enabling column-level metadata, which collectively enhance both I/O performance and overall efficiency.

Moreover, the complexity associated with managing indexes and maintaining data integrity in traditional database management systems can become a considerable barrier when dealing with large-scale or evolving datasets. ParquetDB offers a more streamlined approach to data management, removing the need for complex index maintenance while still enabling efficient querying through PyArrow integration. The ability to directly work with data in a columnar format allows for more intuitive data operations, including predicate pushdown and leveraging column statistics, thus reducing unnecessary data loads and enhancing overall query performance.

The adaptability of ParquetDB also represents a significant advancement over conventional systems. Traditional relational databases, while offering strong data consistency and robust querying capabilities, often enforce a rigid schema that complicates adaptation to new data structures or evolving project needs. This rigidity can create friction in research environments that thrive on adaptability and require rapid prototyping to respond to new challenges. ParquetDB, by contrast, provides schema evolution capabilities, allowing for flexible modifications without the need for extensive reworking of existing datasets. This feature is particularly valuable in iterative data analysis scenarios, where evolving insights may necessitate modifications to the data representation.

\section{Acknowledgments}
We thank the Pittsburgh Supercomputer Center (Bridges2) and San Diego Supercomputer Center (Expanse) through allocation DMR140031 from the Advanced Cyberinfrastructure Coordination Ecosystem: Services \& Support (ACCESS) program, which is supported by National Science Foundation grants \#2138259, \#2138286, \#2138307, \#2137603, and \#2138296. 
We gratefully acknowledge the computational resources provided by the WVU Research Computing Dolly Sods HPC cluster, partially funded by NSF OAC-2117575. Additionally, we recognize the support from the West Virginia Higher Education Policy Commission through the Research Challenge Grant Program 2022 (Award RCG 23-007), as well as NASA EPSCoR (Award 80NSSC22M0173), for their contributions to this work. The work of E.R.H.  is supported by MCIN/AEI/ 10.13039/501100011033/FEDER, UE through projects PID2022-139776NB-C66. K.C. thanks funding from the CHIPS Metrology Program, part of CHIPS for America, National Institute of Standards and Technology, U.S. Department of Commerce.  Certain commercial equipment, instruments, software, or materials are identified in this paper in order to specify the experimental procedure adequately. Such identifications are not intended to imply recommendation or endorsement by NIST, nor are they intended to imply that the materials or equipment identified are necessarily the best available for the purpose.

\section{Appendix 1: Definitions}
\begin{description}

    \item[Schema] A structure that defines the organization, types, and constraints of data within a database or dataset.

    \sloppy
    \item[Schema Alignment] The process of adjusting or reconciling different schemas to ensure data compatibility across systems or datasets. This involves matching fields, data types, and structures to achieve consistency, allowing seamless data integration, transformation, or comparison between sources.

    \item[CSV] Comma-Separated Values, a file format used to store tabular data where each line represents a record and each field is separated by a comma.

    \item[JSON] JavaScript Object Notation, a lightweight format for data exchange that is easy for humans to read and write and easy for machines to parse and generate.

    \item[SQL] Structured Query Language, a standardized language used for managing and querying relational databases. SQL enables users to define, manipulate, and retrieve data with a set structure, allowing for complex queries and operations.

    \item[NoSQL] A category of databases designed to handle unstructured or semi-structured data, offering flexible schema designs and scalability. NoSQL databases are typically used for large-scale data and distributed systems, and include types like document, key-value, column-family, and graph databases.

    \item[Relational Database] A type of database that organizes data into tables (relations) with defined relationships between them. Data is structured using schemas and accessed using SQL, providing strong data integrity and consistency. Examples include MySQL, PostgreSQL, and Oracle Database.

    \item[Document Database] A NoSQL database that stores data in the form of documents, often using JSON or BSON formats. Each document contains semi-structured data, allowing for flexible and dynamic schemas. Document databases are well-suited for applications requiring hierarchical or nested data structures, such as MongoDB or CouchDB.

    \item[ASCII] American Standard Code for Information Interchange, a character encoding standard for electronic communication, representing text in computers and other devices.

    \item[UTF] Unicode Transformation Format, a family of character encoding standards designed to represent all possible characters from all writing systems.

    \item[Random Access Memory (RAM)] A type of computer memory that can be accessed randomly and is used to store working data and machine code currently being used.

    \item[Disk] A data storage device that stores digital information, using either magnetic or solid-state technology.

    \item[Encoding] The process of converting data into a specific format to facilitate storage, transmission, or processing.

    \item[Compression] A technique to reduce the size of data for storage or transmission purposes by removing redundancies.

    \item[Portability] Refers to how data is stored and accessed. Data in database servers is often managed on remote servers, making it less directly accessible from the file system. In contrast, formats like CSV are more portable because they are stored locally as single files, easily moved, and accessed on different systems.

    \item[Schema Evolvability] The ability of a schema to adapt or change over time to accommodate new data structures or fields.

    \item[Data Validation] The process of ensuring that data conforms to specified formats, constraints, and ranges to maintain data quality and reliability.

    \item[Scalability] The capability of a system or process to handle a growing amount of work or to accommodate expansion.

    \item[Batch Support] The ability of a system to process a group of tasks or data entries as a single unit, typically to improve performance or efficiency.
    
    \item[Concurrency] The ability of a database or system to allow multiple operations or transactions to occur simultaneously, which is essential for high-performance applications.

    \item[ACID Compliance] A set of properties that ensure reliable processing in a database. ACID stands for Atomicity, Consistency, Isolation, and Durability, which together guarantee that database transactions are processed accurately and remain consistent, even in cases of failure.

    \item[Predicate Pushdown] An optimization technique used in databases, particularly in columnar storage formats, where filters or conditions (predicates) are applied as early as possible during data retrieval. By pushing down predicates to the storage layer, the database reduces the amount of data read and processed, improving query performance.

    \item[Column Pushdown] An optimization in columnar databases where only the required columns are read from storage rather than the entire dataset. This minimizes data retrieval and processing time, especially in analytics workloads where only specific attributes are needed from large datasets.

\end{description}

\clearpage

\bibliographystyle{unsrt}


\end{document}


\maketitle

\section{Parquet File Layout}

The Parquet file structure is engineered for optimal performance (refer to Figure Y for an overview of the Parquet file layout). Parquet files are partitioned into row groups, which serve as the horizontal segmentation of the dataset. Each row group is further divided into column chunks, and each column chunk comprises pages, representing the smallest storage unit in Parquet. At the conclusion of each Parquet file, a footer is appended, containing essential metadata, including the schema, field-level information, and row group metadata such as the minimum and maximum values for each column within a row group. This footer is fundamental for enabling efficient data access and is integral to the self-descriptive nature of Parquet files. The hierarchical organization of row groups, column chunks, and pages facilitates efficient data compression and retrieval, while also supporting advanced features like predicate pushdown and columnar projection. Parquet files also include metadata at various hierarchical levels: file-level metadata, which details the overarching structure, and row group metadata, which encompasses statistics such as minimum and maximum values that are crucial for query optimization.

\subsection{Pages}

Pages in Parquet files are responsible for storing the actual data, and each page is composed of two principal components: the page header and the data pages. The page header encapsulates metadata pertaining to the page, such as the data type, uncompressed and compressed page sizes, the codec utilized, the data page header, and optionally, the dictionary page header. The data page header conveys details including the number of values, the encoding type, and the definition and repetition level encodings, which facilitate the representation of nested structures. In scenarios employing dictionary encoding, the dictionary page header specifies the count of values within the dictionary. Subsequent to the page header is the dictionary page, if applicable, which contains the dictionary-encoded values. This dictionary page substantially reduces storage requirements by maintaining unique values that are referenced by the data pages. In the absence of dictionary encoding, the data pages directly store the encoded values.

\subsection{Bloom Filter}
This is located before the column index and offset index

\subsection{Page Index}

The Page Index in a Parquet file is located in the footer and provides critical statistics for column data pages, enabling more efficient data scanning. The Page Index contains two main structures: the ColumnIndex, which helps locate data pages containing specific values for a given scan predicate, and the OffsetIndex, which helps navigate by row index to retrieve matching values across different columns within a row group.

These index structures are specifically designed to make data retrieval operations, such as range scans and point lookups, I/O efficient. They allow a reader to access only the data pages necessary for a particular operation, thereby significantly reducing the overhead of reading irrelevant pages. For example, in range scans on sorted columns, the Page Index helps the reader skip non-relevant pages and directly access those with the desired values.

The index structures are stored separately from the row group metadata, right before the footer of the file, ensuring they do not add I/O overhead unless explicitly needed. This separation ensures that readers not requiring selective scans can avoid unnecessary index deserialization costs. The boundaries between pages are recorded using minimum and maximum values for each page, and these boundaries are used by readers to perform efficient searches, such as binary searches for ordered columns or sequential scans for unordered columns.

\subsection{Footer}
The footer in a Parquet file serves as a critical component that consolidates metadata necessary for efficient data operations. It contains comprehensive metadata for the entire file, providing essential information to support efficient data access and processing. The footer includes:

\subsubsection{FileMetaData}
The \texttt{FileMetaData} is the top-level metadata structure in a Parquet file. It encapsulates essential information about the file's schema, version, row groups, and other properties. This metadata enables readers to understand how to interpret and process the data stored within the file. Below is a detailed explanation of each component within the metadata:

\begin{itemize}[label=-]
    \item \textbf{Schema} - The schema describes each column in detail, specifying attributes such as column names, data types, repetition levels, number of child elements, and converted (logical) types.
    \item \textbf{Version} - The version of the Parquet format used to write the file.
    \item \textbf{Number of Rows} - The total number of rows in the file.
    \item \textbf{Key Value Metadata} - Optional user-defined key-value metadata.
    \item \textbf{Created by} - The writer version that created the file. Indicates the name and version of the software that wrote the Parquet file, such as "Apache Parquet-MR version 1.10.1" or "pyarrow 3.0.0".
    \item \textbf{Column Order} - A list of the column orders. Specifies the ordering of columns in the file, which can impact how data is read and optimized during query execution.
    \item \textbf{Encryption Algorithm} - Specifies the algorithm used to encrypt the file.
    \item \textbf{Footer Signing Key Metadata} - Metadata related to the key used to sign the footer, for encryption purposes.
\end{itemize}

\subsubsection{Schema}
The schema describes each column in detail, specifying attributes such as column names, data types, repetition levels, number of child elements, and converted (logical) types. The order of columns is important as it affects how data is read and optimized during query execution. The schema provides a detailed blueprint of the data structure within the Parquet file. It includes:

\begin{itemize}[label=-]
    \item \textbf{Column Names} - Identifiers for each field or column in the dataset.
    \item \textbf{Data Types} - Physical data types like \texttt{INT32}, \texttt{FLOAT}, \texttt{BYTE\_ARRAY}, etc.
    \item \textbf{Logical Types (Converted Types)} - Higher-level data types like \texttt{DECIMAL}, \texttt{DATE}, \texttt{TIMESTAMP}, which provide semantic meaning over physical types.
    \item \textbf{Repetition Levels} - Indicate whether fields are required, optional, or repeated (allowing for lists or arrays).
    \item \textbf{Nested Structures} - Information about complex types such as structs, maps, and lists, including the number of child elements and their respective schemas.
\end{itemize}

\subsubsection{Row Group MetaData}
Each row group in a Parquet file represents a horizontal partition of the data, containing a subset of rows for all columns. The metadata for each row group provides crucial information that enables efficient data access, storage optimization, and query execution. Below is a detailed explanation of each component within the row group metadata:

\begin{itemize}[label=-]
    \item \textbf{Column Chunks} - A list of Column Chunk Metadata for each column in the row group. Each column chunk corresponds to the data of a single column within the row group. The column chunk metadata includes information such as the data type, encoding schemes, compression codec, file offsets, and statistics like min and max values. This metadata is essential for readers to locate and interpret the column data correctly, allowing for efficient columnar access and operations like predicate pushdown and data skipping.
    \item \textbf{Total uncompressed byte size} - The total size in bytes of the row group before compression. This value represents the sum of the uncompressed sizes of all column chunks within the row group. It provides an estimate of the raw data size, which can be useful for understanding the level of compression achieved and for planning memory allocation during data processing.
    \item \textbf{Total compressed byte size} - The total size in bytes of the row group after compression. This is the actual size of the row group as stored on disk, after applying compression algorithms. It reflects the storage space utilized by the row group and is critical for I/O operations, as it influences the amount of data that needs to be read from or written to disk.
    \item \textbf{Number of rows} - The total number of rows contained in the row group. This indicates how many rows are present in the row group, which helps in dividing the dataset into manageable chunks for parallel processing. Knowing the number of rows is also important for query planning and optimization, as it affects operations like joins, aggregations, and limit clauses.
    \item \textbf{Sorting columns} - The columns by which the rows in the row group are sorted, including sorting order (ascending or descending). If the data within the row group is sorted based on one or more columns, this metadata specifies those columns and the sort order. Sorted data can significantly enhance query performance by enabling faster data retrieval, efficient range scans, and improved compression ratios. Query engines can leverage this information to optimize execution plans, especially for queries involving order-dependent operations or filters on sorted columns.
    \item \textbf{File Offset} - The byte offset in the file where the row group starts. This offset points to the exact location in the Parquet file where the row group's data begins. It allows readers to seek directly to the row group without scanning the entire file, facilitating random access and efficient data retrieval. This is particularly important in distributed storage systems or when dealing with large files.
    \item \textbf{Ordinal} - The position or index of the row group within the file. The ordinal is a zero-based index indicating the sequence of the row group in the file. It helps in identifying the order of row groups, which can be useful for certain operations like consistent data shuffling, partitioning, or when reconstructing the dataset's original sequence. The ordinal can also assist in correlating row groups across different files or datasets when performing distributed processing.
\end{itemize}

\subsubsection{ColumnChunk MetaData}
Contains metadata specific to the column chunk, such as encoding types, compression codec, data type, number of values, and statistics like min and max values. This is essential for reading and interpreting the column data correctly.

\begin{itemize}[label=-]
    \item \textbf{File Path} - Specifies the relative path to the file where the column chunk is stored. If the column chunk is in the same file as the metadata (which is common), this field may be null or omitted.
    \item \textbf{File Offset} - The byte offset within the file where the column chunk data begins. This allows readers to locate the exact position of the column data in the file.
    \item \textbf{Offset Index Offset} - The byte offset to the offset index for the column chunk. The offset index contains information about the locations of individual pages within the column chunk, which can be used for efficient data access.
    \item \textbf{Offset Index Length} - The length in bytes of the offset index. This tells readers how much data to read starting from the Offset Index Offset to obtain the full offset index.
    \item \textbf{Column Index Offset} - The byte offset to the column index for the column chunk. The column index provides min and max statistics for each page within the column chunk, facilitating faster queries by enabling data skipping.
    \item \textbf{Column Index Length} - The length in bytes of the column index. This indicates how much data to read from the Column Index Offset to retrieve the entire column index.
    \item \textbf{Crypto MetaData} - Contains information related to encryption, such as the encryption algorithm used and key metadata. This is crucial for decrypting the column chunk if encryption is applied.
    \item \textbf{Encrypted Metadata} - If the column metadata itself is encrypted, this field contains the encrypted bytes. This ensures that sensitive metadata is protected, and only authorized readers with the correct decryption keys can access it.
\end{itemize}

\subsubsection{Column MetaData}
The metadata for a column chunk provides detailed information about how the column's data is stored, encoded, and compressed within a Parquet file. This metadata is crucial for correctly reading and interpreting the column data, optimizing query performance, and managing resources efficiently. Below is an in-depth explanation of each component within the

\begin{itemize}[label=-]
    \item \textbf{Type} - The data type of the column. Specifies the physical data type used to store the column's values in the Parquet file.
    \item \textbf{Encodings} - A list of the encoding types used for the column data. Lists all encoding mechanisms applied to the column's data pages. Encodings optimize storage and speed up data processing by reducing data size and enabling efficient decompression.
    \item \textbf{Path in the schema} - The hierarchical path of the column in the schema. Represents the nested path to the column within the Parquet schema, especially important for complex data structures like nested records or arrays. For example, a path might be \texttt{["customer", "address", "city"]} for a nested field. This helps map the column data to the correct field in the application's data model.
    \item \textbf{Codec} - The compression codec used to compress the column data. Specifies the compression algorithm applied to the column's data pages. Compression reduces disk space usage and I/O costs.
    \item \textbf{Number of values} - The total number of values in the column chunk, including nulls. Represents the total count of logical values stored in the column chunk. This includes both null and non-null values.
    \item \textbf{Total uncompressed size} - The total size in bytes of the column chunk data before compression. Indicates the amount of data before any compression is applied.
    \item \textbf{Total compressed size} - The total size in bytes of the column chunk data after compression. The actual size of the data stored on disk. This impacts storage and retrieval performance.
    \item \textbf{Key\_value\_metadata} - Optional key-value metadata specific to the column. Allows for custom metadata to be attached to the column.
    \item \textbf{Data page offset} - The byte offset in the file to the first data page of the column chunk. Specifies where the column's data pages begin in the file. Data pages contain the actual encoded and compressed data.
    \item \textbf{Index page offset} - The byte offset in the file to the index page for the column chunk. If present, the index page contains information that allows for faster data access.
    \item \textbf{Dictionary\_page\_offset} - The byte offset in the file to the dictionary page of the column chunk. Relevant when dictionary encoding is used. The dictionary page contains the unique values (dictionary) that data pages reference.
    \item \textbf{Statistics} - Statistical information for the column, including min and max values. Provides aggregate statistics about the column data, such as:
    \begin{itemize}
        \item \textbf{Minimum Value}: The smallest value in the column chunk.
        \item \textbf{Maximum Value}: The largest value in the column chunk.
        \item \textbf{Null Count}: The number of null values.
        \item \textbf{Distinct Count}: The number of distinct values (optional).
        \item \textbf{Sum}: The total sum of all values (optional).
    \end{itemize}
    \item \textbf{Encoding\_stats} - Statistics about the encodings used in the column chunk, including counts of pages encoded with each encoding type. Details how different encoding methods are applied across the data pages in the column chunk.
    \item \textbf{Bloom\_filter\_offset} - The byte offset in the file to the Bloom filter data for the column. Points to the location of the Bloom filter, a probabilistic data structure used to test whether an element is a member of a set.
    \item \textbf{Bloom\_filter\_length} - The length in bytes of the Bloom filter data for the column. Indicates the size of the Bloom filter.
\end{itemize}

\section{How Parquet Files are written}
To gain a deeper understanding of how pages are formed within Parquet files, we must explore the underlying mechanisms involved in the process of writing Parquet files (for more detailed information, refer to this article). Although Parquet is a columnar format, its internal representation necessitates writing data row by row. Each row is decomposed into individual columns, and the values of these columns are subsequently added to their respective in-memory column stores. During this phase, metadata such as minimum and maximum value statistics, as well as the count of NULL values, is also updated for each column. At this stage, all data remains in memory.

Once the initial 100 values for a column have been written (equivalent to 100 rows), the Parquet writer evaluates whether the column content exceeds the specified page size threshold, which is typically set at 1 MB. If the raw data size remains below this threshold, subsequent page size checks are dynamically adjusted based on the actual size of the column, rather than occurring after each value or strictly every 100 values. Consequently, the page size limit is not rigidly enforced. Should the raw data size exceed the page size threshold, the column content is compressed (if compression is enabled for the Parquet file) and then flushed into the page store. Each page also contains metadata, referred to as the page header, which includes the uncompressed and compressed sizes, the number of values, and statistics such as the minimum and maximum values for the column, along with the count of NULL values.

After writing the first 100 rows to memory, the Parquet writer checks whether the data size exceeds the specified row group size (block size), which by default is 128 MB. This size includes both the uncompressed size of the data in the column store (not yet flushed to the page store) and the compressed data already held in the page store for each column.

If the data size does not exceed the specified row group size, the Parquet writer estimates the next size check based on the average row size. This next check may occur after 100 rows or even after 10,000 rows, indicating that the row group size limit is not strictly enforced.

If the data size exceeds the specified row group size, the Parquet writer flushes the content of all column stores into their respective page stores and subsequently flushes all page stores to the output stream, column by column. This is the first instance where data is written to an external stream (HadoopPositionOutputStream), thereby making it potentially visible to external components, such as S3 Multipart Upload transfer threads, which may start uploading data to S3 in the background.

Once the row group has been flushed, the memory occupied by the current column and page stores is freed, although garbage collection cycles may be required to fully reclaim this memory.

It is important to note that the row group content itself does not include metadata (e.g., statistics or offsets). Instead, row group metadata is appended to the Parquet file footer.

After all row groups have been written and before the file is closed, the Parquet writer appends a footer to the end of the file. This footer includes the file schema (column names and their respective data types), along with detailed information for each row group, such as the total size, number of rows, and column-specific statistics (e.g., minimum and maximum values, and the count of NULL values). These statistics are provided for each row group individually, rather than for the entire file.

Storing metadata in the footer allows the Parquet writer to avoid keeping the entire file in memory or on local disk, thereby allowing row groups to be safely flushed once they are completed.